\documentclass[acmtog,nonacm,balance=True]{acmart}
\renewcommand\footnotetextcopyrightpermission[1]{}
\renewcommand\footnotetextauthorsaddresses[1]{}

\usepackage{booktabs} 

\usepackage{bm}
\usepackage{makecell}
\usepackage{xspace}
\usepackage{enumitem}

\usepackage{multirow}
\AtEndPreamble{
    \usepackage[capitalize]{cleveref}
    \crefname{section}{Sec.}{Secs.}
    \Crefname{section}{Section}{Sections}
    \Crefname{table}{Table}{Tables}
    \crefname{table}{Tab.}{Tabs.}
}

\def\eg{e.g.~}

\usepackage[ruled]{algorithm2e} 

\begin{document}
\title{Neural BRDF Importance Sampling by Reparameterization}

\author{Liwen Wu}
\orcid{0009-0007-2773-2032}
\affiliation{
    \institution{University of California San Diego}
    \country{USA}
}
\email{liw026@ucsd.edu}
\author{Sai Bi}
\orcid{0000-0002-0311-2521}
\affiliation{
    \institution{Adobe Research}
    \country{USA}
}
\email{sbi@adobe.com}
\author{Zexiang Xu}
\orcid{0000-0002-8487-8018}
\affiliation{
    \institution{Hillbot}
    \country{USA}
}
\email{zexiangxu@gmail.com}
\author{Hao Tan}
\orcid{0000-0002-9755-9040}
\affiliation{
    \institution{Adobe Research}
    \country{USA}
}
\email{hatan@adobe.com}
\author{Kai Zhang}
\orcid{0000-0002-1727-1689}
\affiliation{
    \institution{Adobe Research}
    \country{USA}
}
\email{kaiz@adobe.com}
\author{Fujun Luan}
\orcid{0000-0001-5926-6266}
\affiliation{
    \institution{Adobe Research}
    \country{USA}
}
\email{fluan@adobe.com}
\author{Haolin Lu}
\orcid{0009-0008-2595-2493}
\affiliation{
    \institution{Max Planck Institute for Informatics}
    \country{Germany}
}
\email{haolinlu@mpi-inf.mpg.de}
\author{Ravi Ramamoorthi}
\orcid{0000-0003-3993-5789}
\affiliation{
    \institution{University of California San Diego}
    \country{USA}
}
\email{ravir@cs.ucsd.edu}


\begin{abstract}
Neural bidirectional reflectance distribution functions (BRDFs) have emerged as popular material representations for enhancing realism in physically-based rendering. Yet their importance sampling remains a significant challenge.
In this paper, we introduce a reparameterization-based formulation of neural BRDF importance sampling that seamlessly integrates into the standard rendering pipeline with precise generation of BRDF samples.
The reparameterization-based formulation transfers the distribution learning task to a problem of identifying BRDF integral substitutions.
In contrast to previous methods that rely on invertible networks and multi-step inference to reconstruct BRDF distributions, our model removes these constraints,
which offers greater flexibility and efficiency.
Our variance and performance analysis demonstrates that our reparameterization method achieves the best variance reduction in neural BRDF renderings while maintaining high inference speeds compared to existing baselines.
\end{abstract}

%
%
%

%
%

\begin{teaserfigure}
\centering
    \setlength\tabcolsep{0.0pt}
    \includegraphics[width=0.99\linewidth]{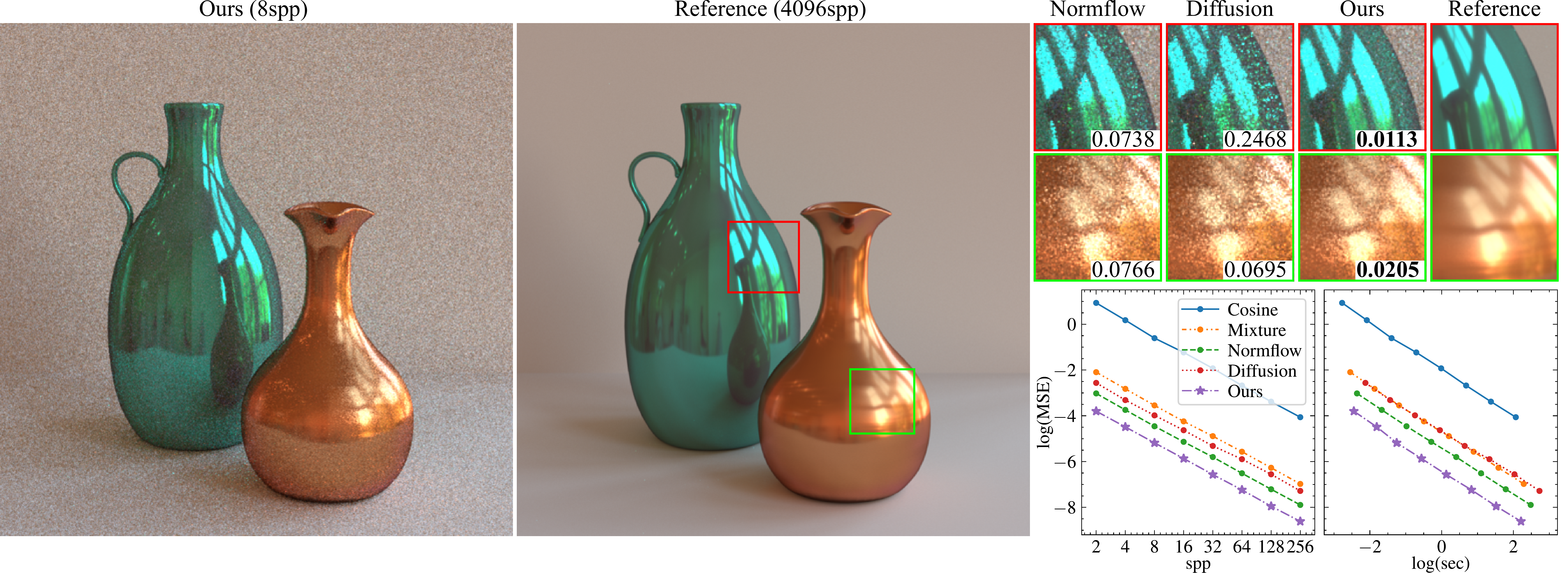}
\vspace{-4pt}
    \caption{
\textbf{Our reparameterization model vs. the baselines.}
Flow-based models~\cite{xu2023neusample,fu2024importance} utilize invertible neural networks for importance sampling by matching the network-induced pdf to the BRDF distribution.
This fails to capture sharp distribution details from the neural BRDFs on the two vases,
leading to noisy renderings of surface reflections (insets in column 3, 4).
Instead, we learn to reparameterize the BRDF (rendering) integral to an easy-to-sample form, which can utilize more flexible neural networks.
The MSE-spp log-log plot (with respect to the full image) suggests our rendering has 2-5$\times$ less variance under same spp (inset in 5th column; numbers show MSE),
and the inference takes 1-1.6$\times$ less time (MSE-sec log-log plot).
The variance from incident radiance dominates the diffuse surface rendering (floor and wall in 1st image) which cannot be effectively reduced by BRDF sampling.
}
    \label{fig:1-teaser}
\end{teaserfigure}
\maketitle

\section{Introduction}
\label{sec:introduction}
Rendering photorealistic images of 3D scenes with physically-based methods requires accurate real-world materials.
Realistic materials often exhibit complex appearances that are difficult to model using standard analytical bidirectional reflectance distribution functions (BRDFs).
Neural networks have shown promise in parameterizing BRDFs to handle such complexity~\cite{sztrajman2021neural,kuznetsov2021neumip,baatz2022nerf,zeltner2024real,dou2024real,fan2023neural}. 
However, importance sampling neural BRDFs for variance reduction in Monte Carlo rendering remains challenging.
In this paper, we present a novel approach to frame importance sampling as learning a reparameterization of the BRDF,
enabling both high-quality sampling and computational efficiency.

Previous works learn to sample BRDFs using flow-based generative models (normalizing flow~\cite{xu2023neusample,muller2019neural} and diffusion models~\cite{fu2024importance}).
These models construct invertible transformations between an easy-to-sample prior distribution and the target BRDF distribution by matching the underlying probability density function (pdf).
The invertibility allows easy derivation of the pdf required for model training and unbiased Monte Carlo rendering.
However, it imposes multi-step transformations, whose computation can be a bottleneck during rendering.

Observing that the probability transform is essentially a substitution rule for pdf integration,
we reinterpret the distribution matching objective as learning a reparameterization of the pdf integral to align with the prior distribution (\cref{subsec:reparam}).
This integral matching objective eliminates the need for invertibility.
Embedding the change-of-variable rule into a neural network,
our method enables efficient single-step BRDF sampling with flexible representations while maintaining unbiased Monte Carlo rendering through the change-of-variable theorem.

Although our model does not explicitly provide its pdf, we demonstrate in \cref{subsec:mis} that it can still support multiple importance sampling (MIS)~\cite{veach1995optimally} using an approximated pdf. This approximation does not introduce bias, as it is used solely for calculating the MIS weight. Furthermore, it is implemented through a lightweight network, ensuring minimal computational overhead.

Since BRDF distributions are often smooth, 
they can also be reasonably approximated by mixtures of analytical distributions~\cite{zeltner2024real,sztrajman2021neural,xu2023neusample,fan2022neural},
which are efficient to sample but lack flexibility in modeling distribution details.
In contrast, our method achieves similar sampling efficiency with accuracy similar to flow-based models.

As illustrated in \cref{fig:1-teaser}, our reparameterization model achieves the best variance-speed trade-off for rendering with neural BRDFs.
It especially outperforms baseline models on specular materials (2.32-7.14$\times$ faster than analytic mixtures, 1.39-3.77$\times$ faster than flow-based models for equal variance rendering) and is still 1-1.47$\times$ more efficient on less specular materials (\cref{subsec:results}).
In summary, our contributions include:
\begin{itemize}
    \item A novel reparameterization model for learning accurate and efficient BRDF sampling in a single step without needing network invertibility.  
    \item An unbiased method for integrating neural BRDF samplers into the MIS framework without explicit pdf evaluations.
\end{itemize}

\section{Background and Related Work}
\label{sec:related-work}
\noindentparagraph{Neural BRDF.}
A bidirectional reflectance distribution function (BRDF) $f(\mathbf{x},\bm{\omega}_i,\bm{\omega}_o\!)$ describes the light transport between incident (light) direction $\bm{\omega}_i$ and outgoing direction $\bm{\omega}_o$ at surface location $\mathbf{x}$.
We take the projected hemisphere ($\mathbb{D}^2$) parameterization~\cite{xu2023neusample} of $\bm{\omega}_i$ and $\bm{\omega}_o$ in this paper.
Traditional BRDF representations rely on analytic microfacet models~\cite{walter2007microfacet,cook1982reflectance} or tabulated bidirectional texture functions (BTF)~\cite{dana1999reflectance},
which suffer from either inadequate representation power for complex real-world materials or heavy storage cost.
To overcome these limitations, early works of \citet{rainer2019neural,rainer2020unified} leveraged neural networks to compress BTF data,
while \citet{kuznetsov2021neumip} directly fit material data using coordinate-based multi-layer perceptrons (MLP).
The coordinate-MLP-based representation effectively reconstructs complex appearance features from micro-geometry, shadows, inter-reflection effects,
and has become popular in the follow-up works with focus on: enhanced meso-geometry modeling~\cite{kuznetsov2022rendering,diolatzis2023mesogan,baatz2022nerf};
improved parameterization and layering~\cite{zeltner2024real,sztrajman2021neural,fan2022neural};
and better input encoding strategies~\cite{fan2023neural,dou2024real}. 
\citet{tg2024neupress,tg2024neural} extend neural representations to include subsurface scattering effects, 
and \citet{mullia2024rna} combine physically-based rendering with neural radiance fields~\cite{mildenhall2020nerf} to create renderable 3D assets.

\noindentparagraph{BRDF importance sampling} aims to approximate (the luminance of) $f$ by a conditional distribution $p(\bm{\omega}_i|\bm{\omega}_o,\mathbf{x})\approx f(\mathbf{x},\bm{\omega}_i,\bm{\omega}_o)/F$ with $F=\int\!f\mathrm{d}\bm{\omega}_i$ that can be sampled to build a low-variance Monte Carlo estimator for the rendering equation~\cite{kajiya1986rendering}:
\begin{align}
L_o(\mathbf{x},\bm{\omega}_o)&=\int L_i(\mathbf{x},\bm{\omega}_i)f(\mathbf{x},\bm{\omega}_i,\bm{\omega}_o)\mathrm{d}\bm{\omega}_i
\label{eq:2-bsdfsampling}\\
&=\mathop{\mathbb{E}}_{\bm{\omega}_i\sim p}\!\left[\frac{L_i(\mathbf{x},\bm{\omega}_i)f(\mathbf{x},\bm{\omega}_i,\bm{\omega}_o)}{p(\bm{\omega}_i|\bm{\omega}_o,\!\mathbf{x})}\right].
\label{eq:2-estimator}
\end{align}%
$L_o(\mathbf{x},\bm{\omega}_o),L_i(\mathbf{x},\bm{\omega}_i)$ are outgoing and incident radiance (recursively defined by \eg path tracing).
For analytic BRDFs, inverse transform sampling is commonly employed~\cite{walter2007microfacet,heitz2018sampling,dupuy2023sampling},
while tabulation methods~\cite{lawrence2004efficient,dupuy2018adaptive} can be used to sample arbitrary distributions at the cost of high memory usage.
In the context of neural BRDFs, more general sampling methods utilize analytic mixtures and flow-based generative models trained by minimizing the forward KL-divergence:
\begin{equation}
    \mathcal{L}_\text{KL} = \mathop{\mathbb{E}}_{\bm{\omega}_i\sim f/F}
    \left[
    \log\left(\frac{f(\mathbf{x},\bm{\omega}_i,\bm{\omega}_o)/F}{p(\bm{\omega}_i|\bm{\omega}_o,\mathbf{x})}\right)
    \right]
\end{equation}%
or score/flow matching~\cite{ho2020denoising,song2021score,song2019generative,lipman2022flow}.
These approaches will be the focus of our discussion below.
Additionally, \citet{xu2023neusample} introduce a mixture of histograms model,
but it relies on intensive usage of histograms, 
resulting in a very large model size that cannot be scaled down for fair comparison (\cref{subsec:baselines}).
\citet{bai2022bsdf} pre-bake optimal transport of discretized BRDF samples into neural networks for sampling but their method is biased.
Therefore, we exclude these two methods from our comparisons.

\noindentparagraph{Analytic mixture models} approximate the BRDF distributions using combinations of easy-to-sample primitive distributions with their parameters predicted by neural networks.
Common primitive distributions include a Lambertian lobe combined with Blinn-Phong models~\cite{sztrajman2021neural}, 2D Gaussians~\cite{xu2023neusample,fan2022neural}, and GGX distributions~\cite{zeltner2024real}.
While analytic mixtures cannot perfectly fit the BRDF distributions,
they offer the advantage of fast sampling and pdf evaluation,
making them suitable for real-time applications~\cite{zeltner2024real}.
They can also be represented without neural networks~\cite{herholz2018unified}.

\noindentparagraph{Flow-based generative models} and our method both construct a probability transform  $\mathbf{T}(\mathbf{z}|\bm{\omega}_o,\mathbf{x})$ that maps the prior-distribution sample $\mathbf{z}\!\sim\!q(\mathbf{z})$ to the target $\bm{\omega}_i$ with the pdf obtained by taking the inverse Jacobian $\mathbf{J}_{\mathbf{T}^{-1}}$:
\begin{gather}
    \bm{\omega}_i\sim p:\;\bm{\omega}_i=\mathbf{T}\left(\mathbf{z}|\bm{\omega}_o,\mathbf{x}\right),\;\mathbf{z}\sim q(\mathbf{z})
    \label{eq:2-flow}
    \\
    p(\bm{\omega}_i|\bm{\omega}_o,\mathbf{x})=q\left(\mathbf{T}^{-1}\left(\bm{\omega}_i|\bm{\omega}_o,\mathbf{x}\right)\right)\left|\det\mathbf{J}_{\mathbf{T}^{-1}}\!\!\left(\bm{\omega}_i|\bm{\omega}_o,\mathbf{x}\right)\right|.
    \label{eq:2-flowpdf}
\end{gather}%
Normalizing flow~\cite{dinh2014nice,rezende2015variational} utilizes a sequence of coupling neural networks to build $\mathbf{T}$,
which transform each dimension of $\mathbf{z}$ alternately through basic invertible functions, such as the affine function~\cite{dinh2022density} used by~\citet{xie2019multiple} and quadratic splines~\cite{muller2019neural,durkan2019neural} used by \citet{xu2023neusample}.
A comprehensive discussion of normalizing flow variants can be found in the survey of~\citet{kobyzev2020normalizing}.
The transformation can also be modeled continuously by evolving a neural ordinary differential equation (ODE)~\cite{chen2018neural}.
This continuous formulation is equivalent to the popular diffusion~\cite{ho2020denoising,song2021score,song2019generative}/flow matching~\cite{lipman2022flow} models,
which can be trained using a simulation-free score/flow matching objective to bypass the expensive ODE simulation needed for optimizing KL-divergence (see~\cite{yang2023diffusion} for a detailed survey on diffusion models).
\citet{fu2024importance} are the first to apply diffusion models for importance sampling measured BRDFs using IADB~\cite{heitz2023iterative} with rectified flow distillation~\cite{liu2022flow} to speed up inference.
Although flow-based models excel at recovering BRDF distribution details,
both couplings and ODE solving require multiple neural network evaluations,
resulting in much slower sampling compared to analytic mixtures.

\noindentparagraph{MIS.}
To achieve optimal variance reduction, 
multiple importance sampling (MIS)~\cite{veach1995optimally} linearly combines BRDF sampling $p$ and emitter sampling $p_e$ with weights $w,w_e,w\!+\!w_e\!=\!1$ ($\bm{\omega}_o,\mathbf{x}$ are omitted here):
\begin{align}
\begin{gathered}
L_o\!=\!\int\!\!w(\bm{\omega}_i)L_i(\bm{\omega}_i)f(\bm{\omega}_i) \mathrm{d}\bm{\omega}_i
+\!\int\!\!w_e(\bm{\omega}_i)L_i(\bm{\omega}_i)f(\bm{\omega}_i)\mathrm{d}\bm{\omega}_i\\
=\!\!\!\mathop{\mathbb{E}}_{\bm{\omega}_i\sim p}\!\left[w(\bm{\omega}_i)\frac{L_i(\bm{\omega}_i)f(\bm{\omega}_i)}{p(\bm{\omega}_i)}\right]
\!+\!\!\mathop{\mathbb{E}}_{\bm{\omega}_i\sim p_e}\!\left[w_e(\bm{\omega}_i)\frac{L_i(\bm{\omega}_i)f(\bm{\omega}_i)}{p_e(\bm{\omega}_i)}\right],
\end{gathered}
\label{eq:2-mis}
\\
w(\bm{\omega}_i)=\frac{p^2(\bm{\omega}_i)}{p^2(\bm{\omega}_i)+p_e^2(\bm{\omega}_i)},
\;
w_e(\bm{\omega}_i)=\frac{p_e^2(\bm{\omega}_i)}{p^2(\bm{\omega}_i)+p_e^2(\bm{\omega}_i)}.
\label{eq:2-misweight}
\end{align}%
We use the power heuristic~\cite{veach1995optimally} in \cref{eq:2-misweight}, while other alternatives~\cite{grittmann2019variance,west2022marginal,west2020continuous} also exist.
Other variance reduction techniques can be used on top of our method and MIS,
such as path guiding~\cite{vorba2014line,muller2017practical,dodik2022path,diolatzis2020practical} that samples the approximated incident radiance rather than the emitter,
product sampling the emitter times the BRDF term~\cite{herholz2016product,hart2020practical,clarberg2008practical}, 
and their realizations by neural networks~\cite{muller2019neural,litalien2024neural,zhu2021hierarchical,zhu2021photon}.
\section{BRDF Sampling by Reparameterization}
\label{sec:method}
Our method is inspired by the fact that the optimal probability transform in \cref{eq:2-flow} is a change-of-variable rule that reparameterizes the BRDF distribution $f/F$ to the prior $q$ ($\bm{\omega}_o,\mathbf{x}$ and sometimes $\bm{\omega}_i,\mathbf{z}$ are omitted for simplicity):
\begin{equation}
\begin{gathered}
\mathbf{T}=\mathop{\text{argmin}}_{\mathbf{T}}
\mathop{\mathbb{E}}_{\mathbf{z}\sim q}\left[
\left(f\left(\mathbf{T}\left(\mathbf{z}\right)\right)\left|\det\mathbf{J}_\mathbf{T}\left(\mathbf{z}\right)\right|/F-
q(\mathbf{z})
\right)^2
\right]\\
\text{s.t. } \bm{\omega}_i=\mathbf{T}(\mathbf{z}) \text{ is a valid reparameterization of } f.
\end{gathered}
\label{eq:3-reparam}
\end{equation}%
At convergence,
$f\!\left(\mathbf{T}\right)\!\left|\det\mathbf{J}_\mathbf{T}\right|\!/\!F\!\approx\!q\!\Rightarrow\!
f\!/\!F\!\approx\!q(\mathbf{T}^{-1})\left|\det\mathbf{J}_{\mathbf{T}^{-1}}\right|\!=\!p\;\forall \bm{\omega}_i$ by the inverse function theorem,
so the above condition is sufficient.
To see it is also necessary,
we substitute $\bm{\omega}_i$ by $\mathbf{T}(\mathbf{z})$ in the BRDF integral by the change of variable theorem:
\begin{align}
    \int\!\!f(\bm{\omega}_i)\mathrm{d}\bm{\omega}_i
=\int\!\!f\!\left(\mathbf{T}\left(\mathbf{z}\right)\right)\left|\det\mathbf{J}_\mathbf{T}\!\left(\mathbf{z}\right)\right|\mathrm{d}\mathbf{z}.
\tag*{}
\end{align}%
The reparameterized integrand indeed equals to $Fq$ when $p\!=\!f/F$:
\begin{align}
f\!\left(\mathbf{T}\left(\mathbf{z}\right)\right)\left|\det\mathbf{J}_\mathbf{T}\!\left(\mathbf{z}\right)\right|
&=f\left(\mathbf{T}\!\left(\mathbf{z}\right)\right)\!
/\left|\det\mathbf{J}_{\mathbf{T}^{-1}}\!\left(\mathbf{T}\left(\mathbf{z}\right)\right)\right|
\tag*{}
\\
&=f\left(\mathbf{T}\!\left(\mathbf{z}\right)\right)q(\mathbf{z})/
p\!\left(\mathbf{T}(\mathbf{z})\right)
\tag{\text{apply \cref{eq:2-flowpdf}}}
\\
&=Fq(\mathbf{z})
\tag{\text{apply } \(p\!=\!f/F\)}.
\end{align}%
Note that optimizing \cref{eq:3-reparam} does not require the inverse $\mathbf{T}^{-1}$,
and the Monte Carlo estimator of \cref{eq:2-bsdfsampling} can also be written in a reparameterized form without $\mathbf{T}^{-1}$:
\begin{gather}
L_o=\!\int\!\!L_i\left(\bm{\omega}_i\right) f\!\left(\bm{\omega}_i\right)\mathrm{d}\bm{\omega}_i
    =\!\int\!\!L_i\left(\mathbf{T}(\mathbf{z})\right)f\!\left(\mathbf{T}\!\left(\mathbf{z}\right)\right)\!\left|\det\mathbf{J}_{\mathbf{T}}\!\left(\mathbf{z}\right)\right|\mathrm{d}\mathbf{z}
\tag*{}
\\
    = \mathop{\mathbb{E}}_{\mathbf{z}\sim q}\left[
    \frac{L_i\!\left(\mathbf{T}(\mathbf{z})\right)f\!\left(\mathbf{T}(\mathbf{z})\right)\left|\det\mathbf{J}_{\mathbf{T}}\!\left(\mathbf{z}\right)\right|}{q(\mathbf{z})}
    \right].
\label{eq:3-rendering}
\end{gather}%
These suggest we can construct $\mathbf{T}$ from nearly arbitrary neural networks,
applying \cref{eq:2-flow} to sample $\bm{\omega}_i$,
and estimating \cref{eq:3-rendering} to get the rendering without any bias.
Unlike flow-based models~\cite{xu2023neusample,fu2024importance} constrained by needing an explicit inverse,
we simply parameterize $\mathbf{T}$ as an MLP that maps $\mathbf{z}$ to $\bm{\omega}_i\in\mathbb{D}^2$ (projected hemisphere) in a single step (\cref{subsec:architecture}),
which is much more flexible as well as faster (\cref{subsec:results}).

We discuss the realization of \cref{eq:3-reparam} as a gradient descent loss in \cref{subsec:reparam},
and the network design for $\mathbf{T}$ to be a valid reparameterization is shown in \cref{subsec:valid-reparam}.
Our network is nontrivial to invert,
which creates a challenge for the pdf evaluation.
Thus, we use another neural network to estimate the pdf and integrate our model into MIS without bias (\cref{subsec:mis}).

\begin{figure}[t]
    \centering
    \setlength\tabcolsep{1.0pt}
    \resizebox{0.99\linewidth}{!}{
    \begin{tabular}{ccc}
    \includegraphics[width=0.34\linewidth]{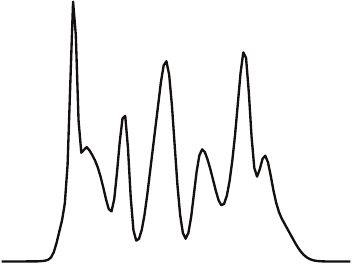}&
    \includegraphics[width=0.28\linewidth]{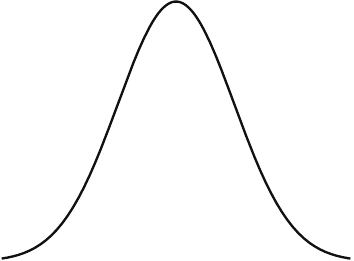}&
    \includegraphics[width=0.28\linewidth]{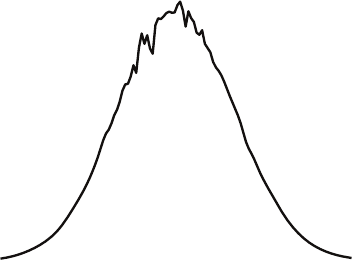}\\
    Target distribution & Prior distribution & Our reparameterization\\
    \includegraphics[width=0.34\linewidth]{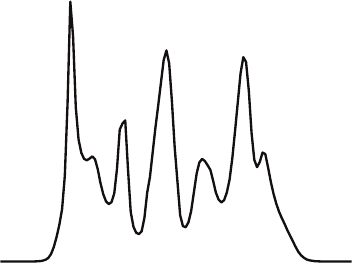}&
    \includegraphics[width=0.34\linewidth]{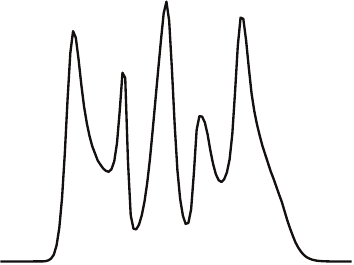}&
    \includegraphics[width=0.34\linewidth]{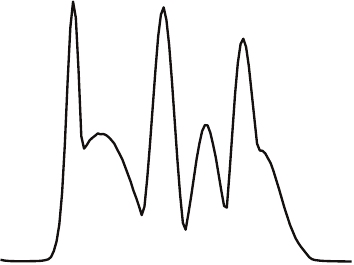}\\
    Our samples & Diffusion samples & Normflow samples
    \end{tabular}
    }
    \caption{\textbf{Toy example in 1D.}
    Our model requires only a small MLP (2 hidden layers, 16 features) to learn the reparameterization of the target distribution to the prior distribution (plot 3 on top), inducing a more accurate importance sampling (plot 1 at bottom) than the baselines (plot 2-3 at bottom) when using a similar network.
    The bottom plots are generated by binning samples into 1D histograms.
    }
    \label{fig:3-reparam}
\end{figure}
\subsection{Matching the reparameterized integral}
\label{subsec:reparam}
The normalization constant $F$ is hard to obtain in practice,
so we match the reparameterized $f/F$ to $q$ by minimizing the negative log likelihood $-\!\!\int\!\!q\log\left(
f(\mathbf{T})\!\left|\det\mathbf{J}_\mathbf{T}\right|\!/\!F
\right)\!\mathrm{d}\mathbf{z}$,
in which $F$ does not contribute to the network gradient:
\begin{align}
-\nabla_\mathbf{T}\!\!\int\!\!q\log\left(
f(\mathbf{T})\!\left|\det\mathbf{J}_\mathbf{T}\right|\!/\!F
\right)\mathrm{d}\mathbf{z}
=&
\tag*{}
\\[-20pt]
-\nabla_\mathbf{T}\!\!\int\!\!q\log\left(
f(\mathbf{T})\!\left|\det\mathbf{J}_\mathbf{T}\right|
\right)&\mathrm{d}\mathbf{z}-\overbrace{\nabla_\mathbf{T}\!\!\int\!\!q\log\left(F\right)\mathrm{d}\mathbf{z}}^{=0}
\tag*{},
\end{align}%
so it can be dropped, resulting in the loss:
\begin{equation}
  \mathcal{L}_\text{nll}=\mathop{\mathbb{E}}_{\mathbf{z}\sim q}\!\left[-\log \left(f(\mathbf{T}(\mathbf{z}))\left|\det\mathbf{J}_{\mathbf{T}}(\mathbf{z})\right|\right)\right].
\label{eq:3-loss}
\end{equation}%
At convergence,
$\mathcal{L}_\text{nll}$ becomes equivalent to the reverse KL-divergence of $p$ and $f/F$ (see supplementary),
which also suggests it learns the correct probability transform.
Figure~\ref{fig:3-reparam} demonstrates a 1D example using a Gaussian and a Gaussian mixture.
It can be seen that optimizing $\mathcal{L}_\text{nll}$ leads to the correct reparameterization that transforms the Gaussian samples to the Gaussian mixture distribution.

\noindentparagraph{Prevent singularity.}
$\mathcal{L}_\text{nll}$ has $\infty$ terms when $f(\bm{\omega}_i)$ or $\left|\det\mathbf{J}_\mathbf{T}(\mathbf{z})\right|$ is zero.
The zero BRDF value is avoided by applying exponential activation to the neural BRDF output or adding a tiny epsilon value.
To prevent zero Jacobian, we split $F$ into two integrals with different reparameterizations:
\begin{equation}
\begin{gathered}
    \int f(\bm{\omega}_i)\mathrm{d}\bm{\omega}_i
    =(1-\alpha)\int f(\bm{\omega}_i)\mathrm{d}\bm{\omega}_i + \alpha \int f(\bm{\omega}_i)\mathrm{d}\bm{\omega}_i\\
    =(1-\alpha)\!\int\!f(\mathbf{T}(\mathbf{z}))|\!\det\mathbf{J}_\mathbf{T}(\mathbf{z})|\mathrm{d}\mathbf{z}
    +\alpha\!\int\!f(\mathbf{I}(\mathbf{z}))|\!\det\mathbf{J}_\mathbf{I}(\mathbf{z})|\mathrm{d}\mathbf{z}\\
    =\int \! \left((1-\alpha)f(\mathbf{T}(\mathbf{z}))|\!\det\mathbf{J}_\mathbf{T}(\mathbf{z})|+\alpha f(\mathbf{I}(\mathbf{z}))|\!\det\mathbf{J}_\mathbf{I}(\mathbf{z})\right) \mathrm{d}\mathbf{z},
\label{eq:3-defensive}
\end{gathered}
\end{equation}%
where $\alpha\!=\!10^{-3}$ and $\mathbf{I}$ covers $\mathbb{D}^2$ with no zero Jacobian:
\begin{equation}
    \mathbf{I}(\mathbf{z})=\frac{\mathbf{z}}{\sqrt{\mathbf{z}\cdot\mathbf{z}+1}},\;
    |\!\det\mathbf{J}_\mathbf{I}(\mathbf{z})|=\frac{1}{(\mathbf{z}\cdot\mathbf{z}+1)^2}.
\end{equation}%
This is similar to defensive sampling where $\mathbf{I}$ is occasionally selected with probability $\alpha$ for sampling to ensure every $\bm{\omega}_i$ can be visited.
The resulting integrand never equals zero,
so it can be stably optimized using negative log likelihood loss:
\begin{equation}
    \mathcal{L}_\text{rep}\!=\!
    \mathop{\mathbb{E}}_{\mathbf{z}\sim q}\!
    \left[
    -\log\left(
    \left(1\!-\!\alpha\right)f\!\left(\mathbf{T}\right)\!
    \left|\det\mathbf{J}_{\mathbf{T}}\right|
    \!+\!
    \alpha f\!\left(\mathbf{I}\right)\!\left|\det \mathbf{J}_\mathbf{I}\right|
    \right)
    \right].
\end{equation}%
In inference, we still only use $\mathbf{T}$ without the defensive sampling.

\begin{figure}[t]
    \centering
    \setlength\tabcolsep{0.4pt}
    \resizebox{0.99\linewidth}{!}{
    \begin{tabular}{cc ccc}
         \includegraphics[width=0.34\linewidth]{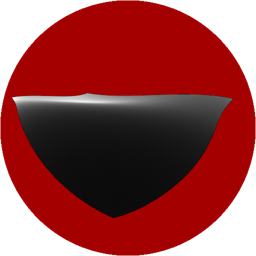}&
         \includegraphics[width=0.34\linewidth]{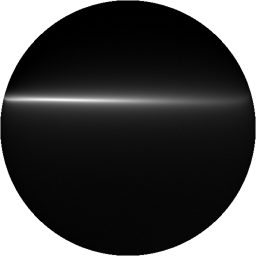}&
         \includegraphics[width=0.34\linewidth]{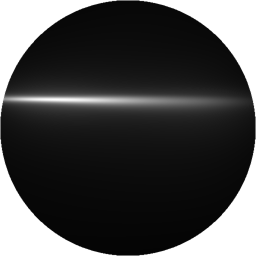}\\
         Uniform & Gaussian & Reference
    \end{tabular}
    }
    \caption{\textbf{Choice of the prior distribution affects the sampling bias.} Images are generated by binning samples to 2D histograms with red regions denoting zero samples.
    When using a distribution with compact support as the prior (left), samples generated by our model may not cover the low probability region of the target distribution (right), resulting in bias.
    The problem can be solved by using a Gaussian-like distribution that has infinite support (middle).
    }
    \label{fig:3-prior}
\end{figure}
\begin{figure}[t]
    \centering
    \setlength\tabcolsep{0.4pt}
    \resizebox{0.99\linewidth}{!}{
    \begin{tabular}{cc ccc}
         \includegraphics[width=0.3\linewidth]{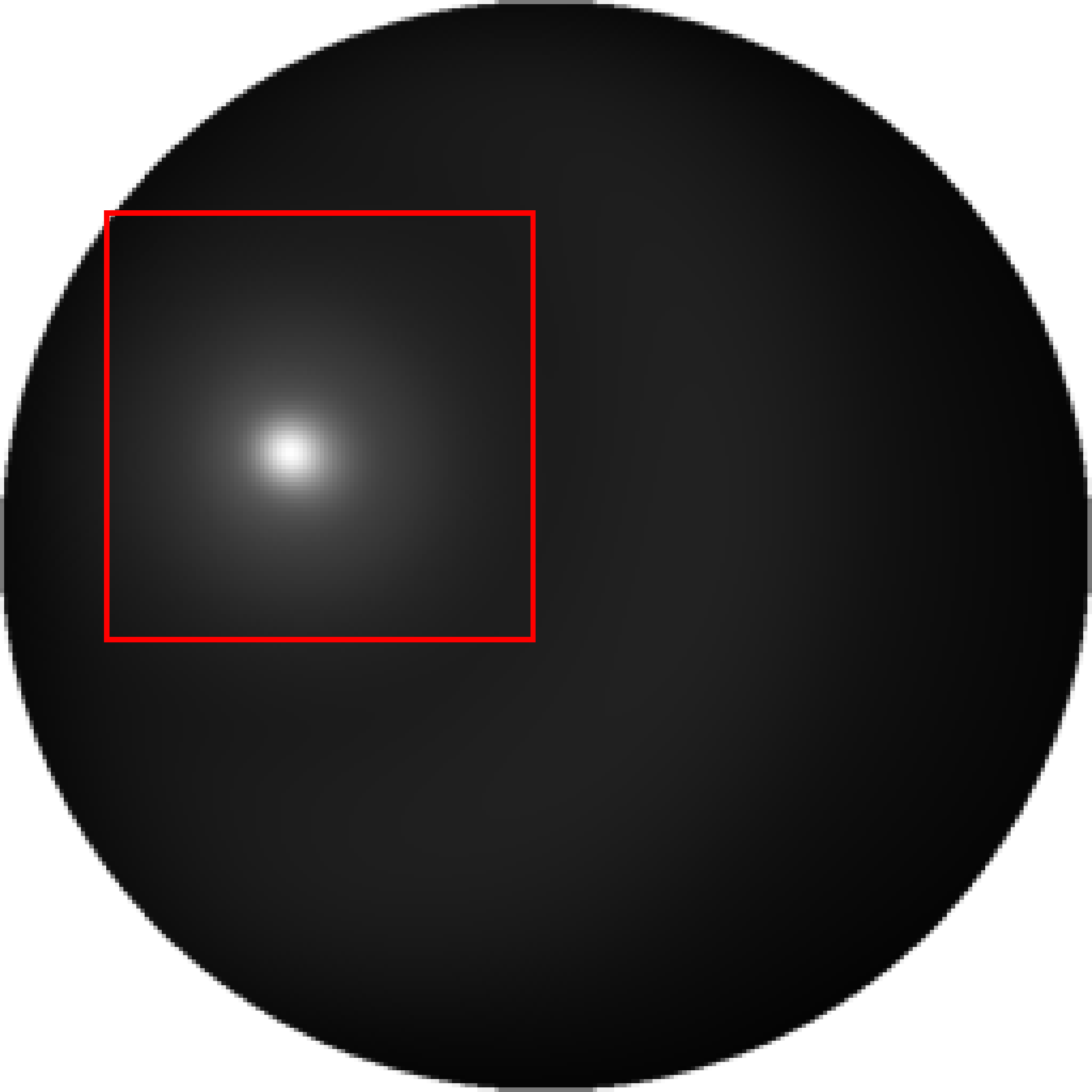}&
         \includegraphics[width=0.3\linewidth]{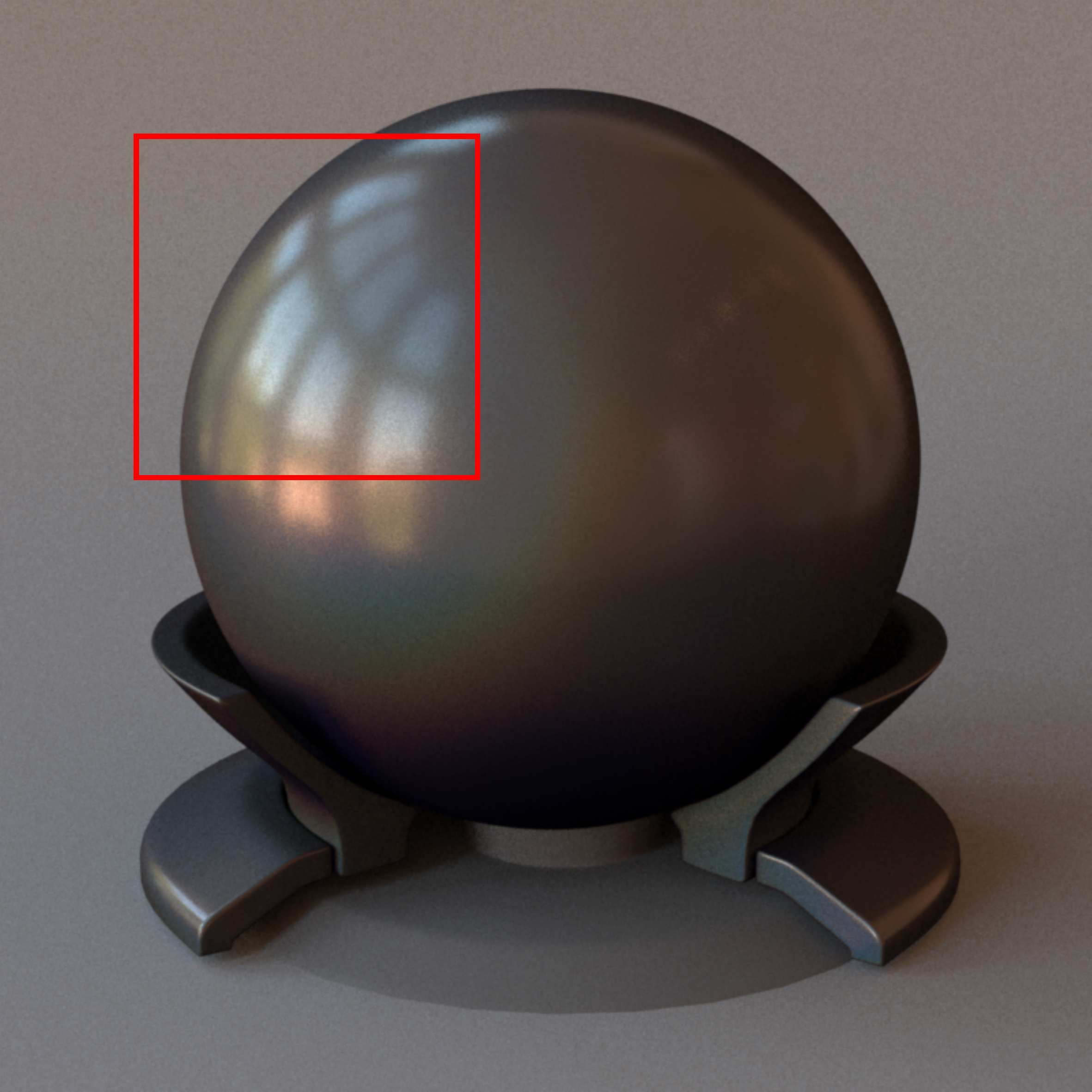}&
         \includegraphics[width=0.15\linewidth]{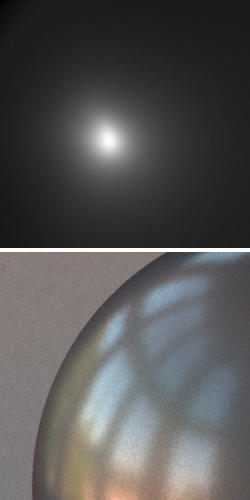}&
         \includegraphics[width=0.15\linewidth]{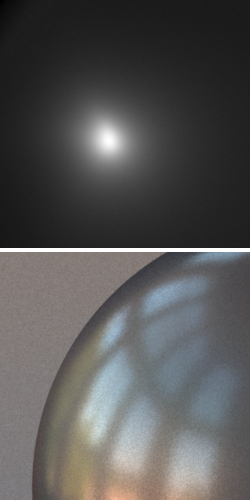}&
         \includegraphics[width=0.15\linewidth]{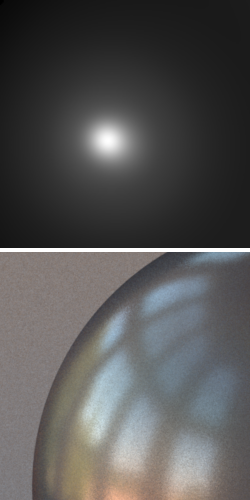}\\
         [-2.5pt]
         \includegraphics[width=0.3\linewidth]{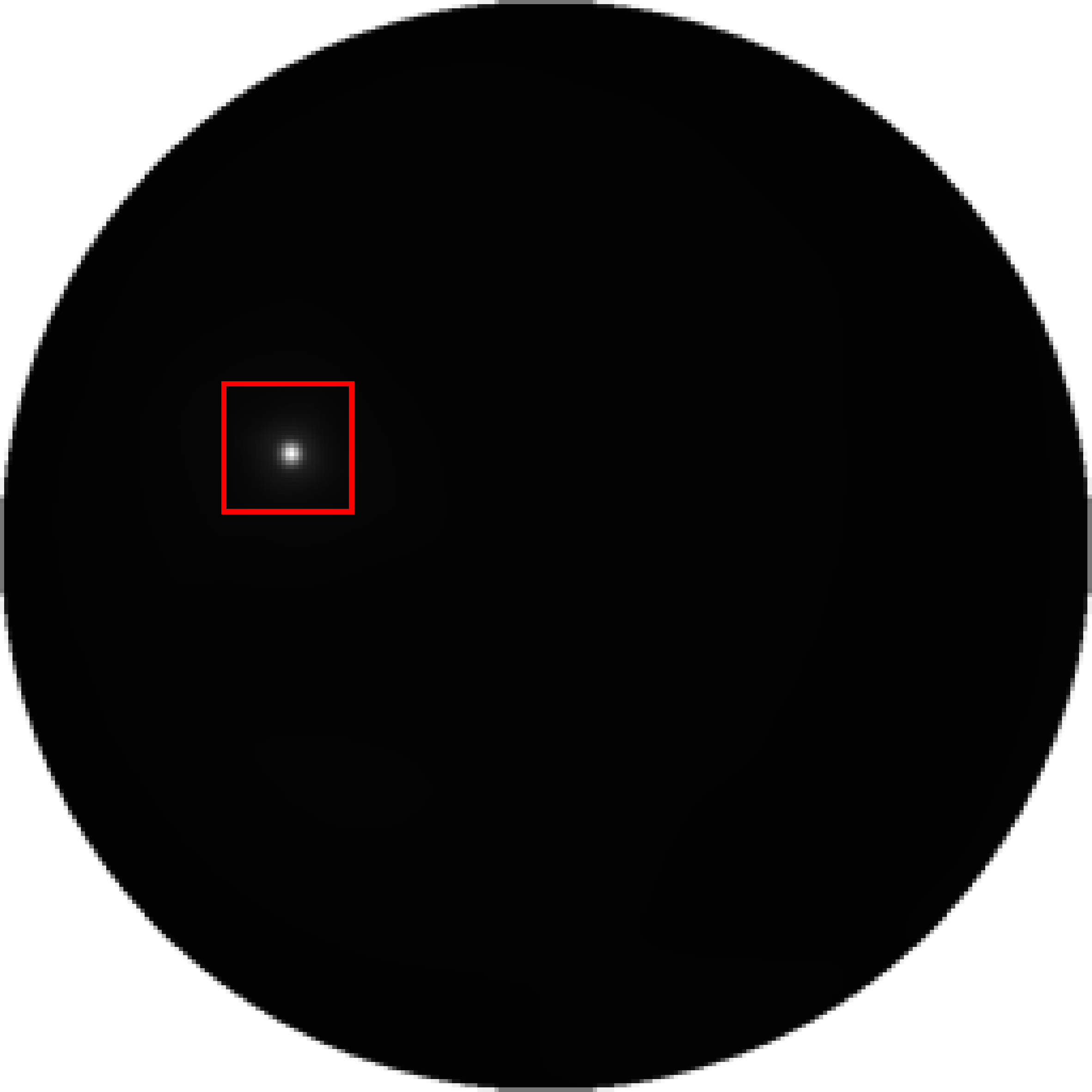}&
         \includegraphics[width=0.3\linewidth]{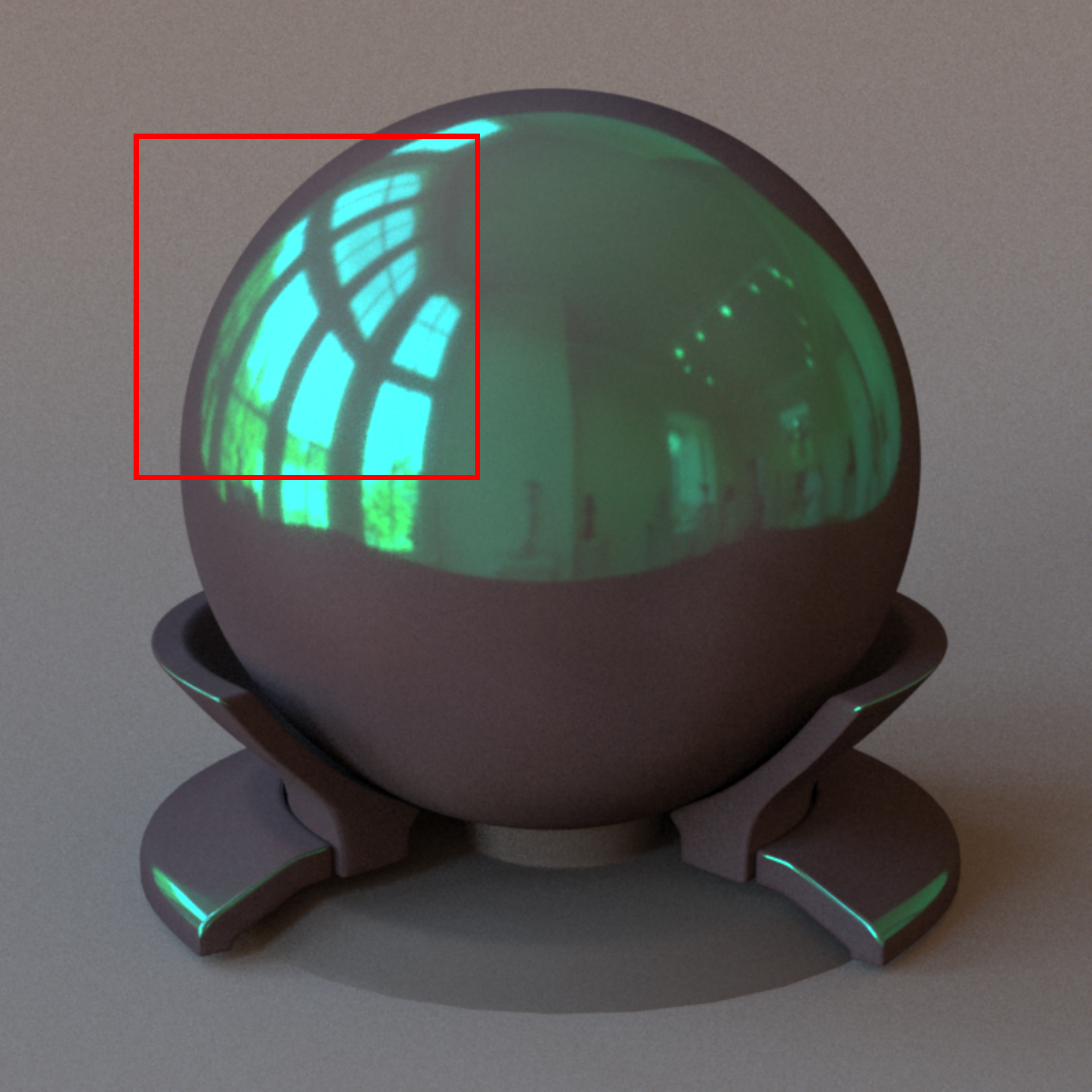}&
         \includegraphics[width=0.15\linewidth]{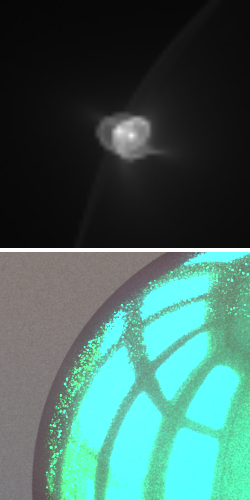}&
         \includegraphics[width=0.15\linewidth]{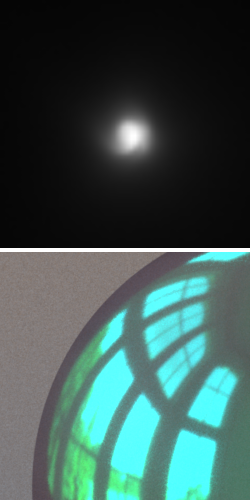}&
         \includegraphics[width=0.15\linewidth]{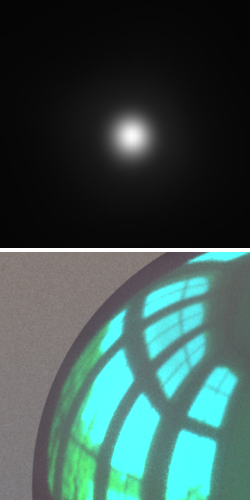}\\
         Pdf & Rendering & $\mathcal{L}_\text{rep}$ & $\mathcal{L}_\text{rep}'$ & Reference
    \end{tabular}
    }
    \caption{\textbf{Comparison between $\mathcal{L}_\text{rep}$ and $\mathcal{L}_\text{rep}'$.}
    Optimizing $\mathcal{L}_\text{rep}$ (column 3) in most cases enables correct reconstructions of BRDF distributions and unbiased renderings (row 1-2).
    But it fails on mirror-like materials (row 3-4).
    This can be fixed by optimizing the upper bound $\mathcal{L}'_\text{rep}$ (column 4).
    The insets show the zoom-ins of reconstructed pdfs (row 1, 3) and the corresponding renderings (row 2, 4).
    }
    \label{fig:3-relaxed}
\end{figure}
\subsection{Ensuring valid reparameterization}
\label{subsec:valid-reparam}
The change of variable theorem requires: 
(1) $\{\mathbf{T}(\mathbf{z}) | q(\mathbf{z})\!>\!0\}\!\supseteq\!\{\bm{\omega}_i | f(\bm{\omega}_i)\!>\!0\}$;
and (2) $\mathbf{T}\!:\!\{\mathbf{z} | q(\mathbf{z})\!>\!0,f(\mathbf{T}(\mathbf{z}))\!\neq\!0\}\!\rightarrow\!\mathbb{D}^2$ is injective.
The first condition is guaranteed for an injective $\mathbf{T}$ when using $q$ with infinite support (Gaussians in our implementation; \cref{fig:3-prior}).
Otherwise, it maps an unbounded domain to a (partially) compact interval of $f(\bm{\omega}_i)\!>\!0$, which leads to zero Jacobians that contradicts with the injectivity by the inverse function theorem.

To ensure the mapping is one-to-one,
we make $\mathbf{T}$ $C^1$-continuous using SiLU activations~\cite{elfwing2018sigmoid,ramachandran2017searching} so that $\det\mathbf{J}_\mathbf{T}$ is $C^0$-continuous.
In this case, $\mathcal{L}_\text{rep}$ at convergence naturally produces an injective map,
or there will be $\det\mathbf{J}_\mathbf{T}=0$ that makes the loss large.
However, this is insufficient if $f(\bm{\omega}_i)\!\approx\!0$ exists,
for which the zero Jacobian does not penalize $\mathcal{L}_\text{rep}$ too much.
Therefore, we further encourage the mapping to only have positive $\det\mathbf{J}_\mathbf{T}$ using an upper bound $\mathcal{L}'_\text{rep}$ (proof in supplementary):
\begin{equation}
\mathcal{L}'_\text{rep}
\!=\!\mathop{\mathbb{E}}_{\mathbf{z}\sim q}\!\left[
-\log\left(
    \left(1\!-\!\alpha\right)\!f\!\left(\mathbf{T}\right)\!
    \max(\det\mathbf{J}_{\mathbf{T}},0)
    \!+\!
    \alpha f\!\left(\mathbf{I}\right)\!\left|\det \mathbf{J}_\mathbf{I}\right|
    \right)
    \right].
\label{eq:3-upper}
\end{equation}%
As shown in \cref{fig:3-relaxed},
using $\mathcal{L}_\text{rep}'$ is essential for samplings of very specular materials to be correct,
where near-zero BRDF responses are common.
An empirical test on the unbiasedness is in \cref{subsec:ablation}.

\subsection{Integrating with MIS}
\label{subsec:mis}
In order to apply MIS with our learned BRDF sampler, $w,w_e$ in \cref{eq:2-misweight} need to be calculated,
which requires pdf evaluation of $\mathbf{T}$.
The exact solution is given by \cref{eq:2-flowpdf}, but finding the inverse of $\mathbf{T}$ is difficult.
However, as long as $w\!+\!w_e\!=\!1$, \cref{eq:2-mis} is still an unbiased estimator,
so we can use an approximated pdf to calculate $w,w_e$.
This is implemented as another MLP $\hat{p}(\bm{\omega}_i)$ trained by the loss:
\begin{equation}
    \mathcal{L}_\text{pdf}
    =\mathop{\mathbb{E}}_{\mathbf{z}\sim q}\left[
    \left(
    \hat{p}(\mathbf{T}(\mathbf{z}))|\!\det\mathbf{J}_\mathbf{T}(\mathbf{z})|-q(\mathbf{z})
    \right)^2
    \right],
\end{equation}%
which gives $\hat{p}(\bm{\omega}_i)\!\approx\!q(\mathbf{T}^{-1}\!(\bm{\omega}_i))|\!\det\mathbf{J}_{\mathbf{T}^{-1}}\!(\bm{\omega}_i)|$ upon $\mathcal{L}_\text{pdf}\approx0$.

\begin{figure}[t]
    \centering
    \setlength\tabcolsep{0.4pt}
    \resizebox{0.99\linewidth}{!}{
    \begin{tabular}{cccc}
         \includegraphics[width=0.28\linewidth]{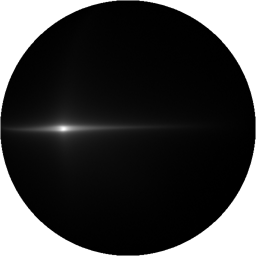}&
         \includegraphics[width=0.28\linewidth]{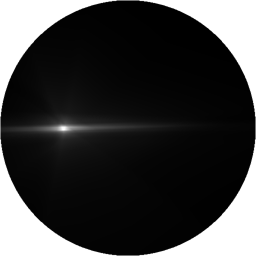}&
         \includegraphics[width=0.28\linewidth]{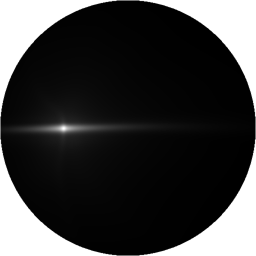}&
         \includegraphics[width=0.28\linewidth]{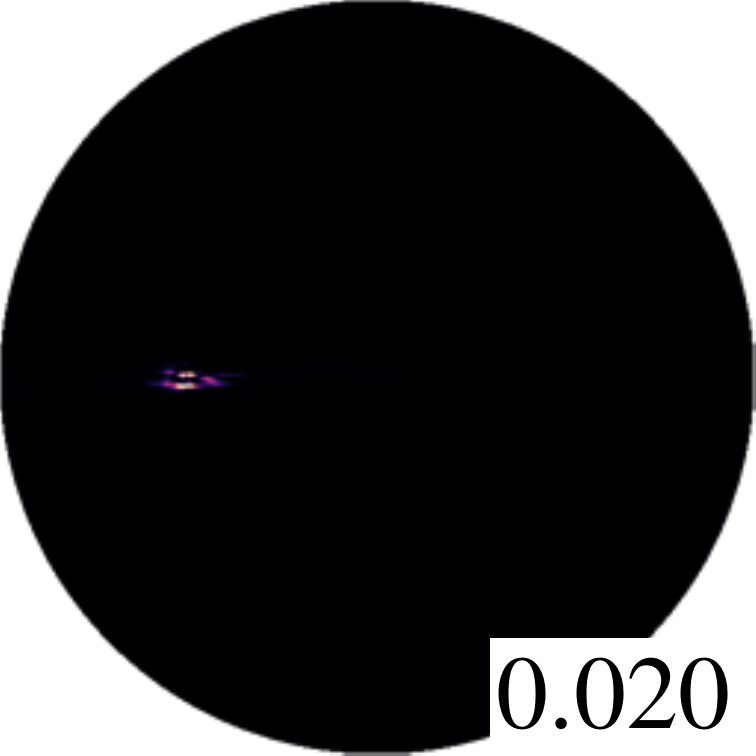}\\[-3.5pt]
         Our sampling & $\hat{p}$ & Reference & Difference$100\times$\\[2.5pt]
         \includegraphics[width=0.28\linewidth]{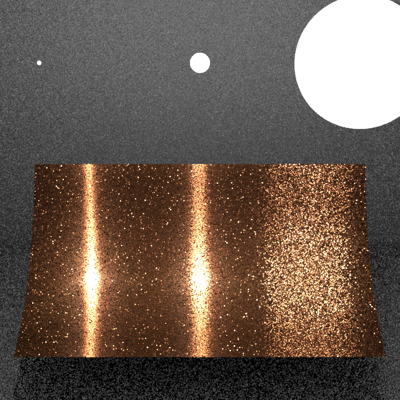}&
         \includegraphics[width=0.28\linewidth]{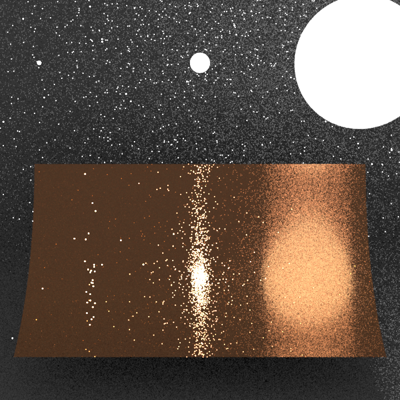}&
         \includegraphics[width=0.28\linewidth]{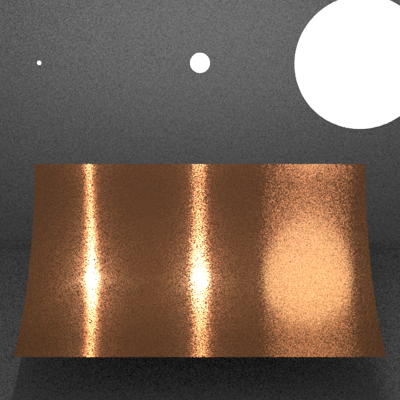}&
         \includegraphics[width=0.28\linewidth]{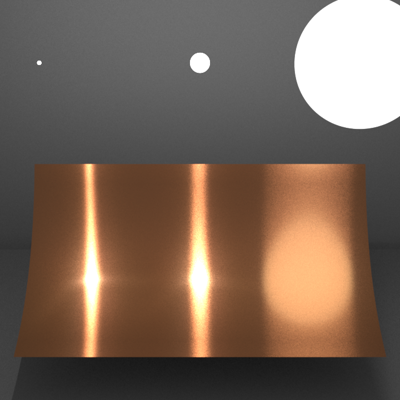}\\[-3.5pt]
         Emitter sampling & BRDF sampling & MIS & Reference
    \end{tabular}
    }
    \caption{\textbf{MIS integrated into our model.}
    The first row shows our pdf approximation $\hat{p}$ for MIS matches our sampling distribution (induced by $\mathbf{T}$) in the first image with little difference (4th image with number showing KL-divergence).
    The second row shows the rendering using emitter sampling (1st image), our BRDF sampling (2nd image), and the combination through MIS (3rd image),
    which is unbiased so matches the reference (4th image).
    }
    \label{fig:3-mis}
\end{figure}

Because the approximated pdf is close to the ground truth, MIS weights given by $\hat{p}$ can also effectively reduce the variance (\cref{fig:3-mis}).
Meanwhile, since solely 
fitting the pdf signal requires no special network design,
$\hat{p}$ can be trained with a smaller network compared to $\mathbf{T}$.
This makes our $w_e$ calculation even faster than other baselines that use the same sampling network for pdf evaluation (\cref{subsec:results}).

\subsection{Network architecture}
\label{subsec:architecture}
\begin{figure}[t]
    \centering
    \setlength\tabcolsep{1.0pt}
    \includegraphics[width=0.99\linewidth]{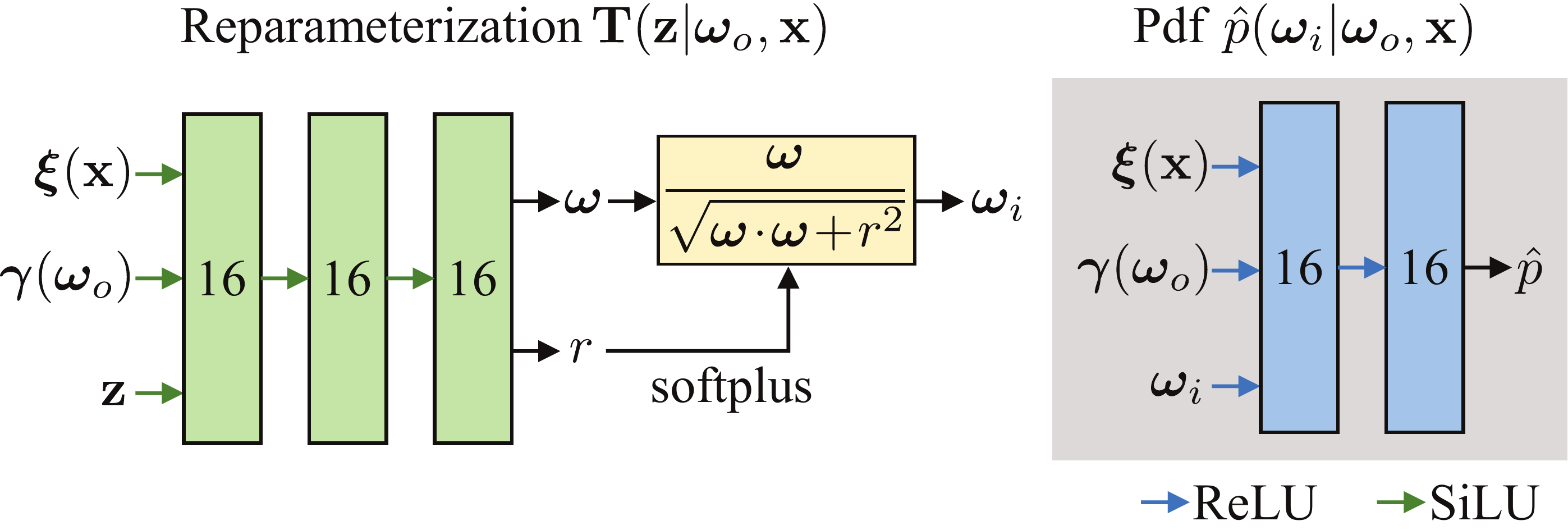}
    \caption{\textbf{Architectures} of reparameterization network (left) and pdf approximation network (right). }
    \label{fig:3-architecture}
\end{figure}

Figure~\ref{fig:3-architecture} shows our network architecture.
$\mathbf{T}$ is a 2-hidden-layer MLP with SiLU activation,
and $\hat{p}$ uses 1 hidden layer with ReLU activation.
We choose the hidden feature size to be 16 to perfectly fit into the CUDA tensor core for fast inference.
The condition $\bm{\omega}_o$ is encoded by positional encoding~\cite{mildenhall2020nerf} $\bm{\gamma}$ of frequency 4,
and the neural texture $\bm{\xi}$ contained in most neural BRDFs~\cite{kuznetsov2021neumip,zeltner2024real,fan2023neural} is reused to encode $\mathbf{x}$.
We do not input $\mathbf{x}$ for non-spatially-varying BRDFs.
Unlike flow-based models that struggle to parameterize the output to $\mathbb{D}^2$,
we do not care about the invertibility so simply let $\mathbf{T}$ produce a 3D vector in the upper-half space and normalize to get $\bm{\omega}_i$.
Compared to directly predicting samples in $\mathbb{R}^2$ domain~\cite{xu2023neusample},
our parameterization prevents invalid samples outside $\mathbb{D}^2$,
and the representation space is more continuous than spherical coordinates~\cite{zhou2019continuity}.

\begin{figure}[t]
    \centering
    \setlength\tabcolsep{1.0pt}
    \includegraphics[width=0.99\linewidth]{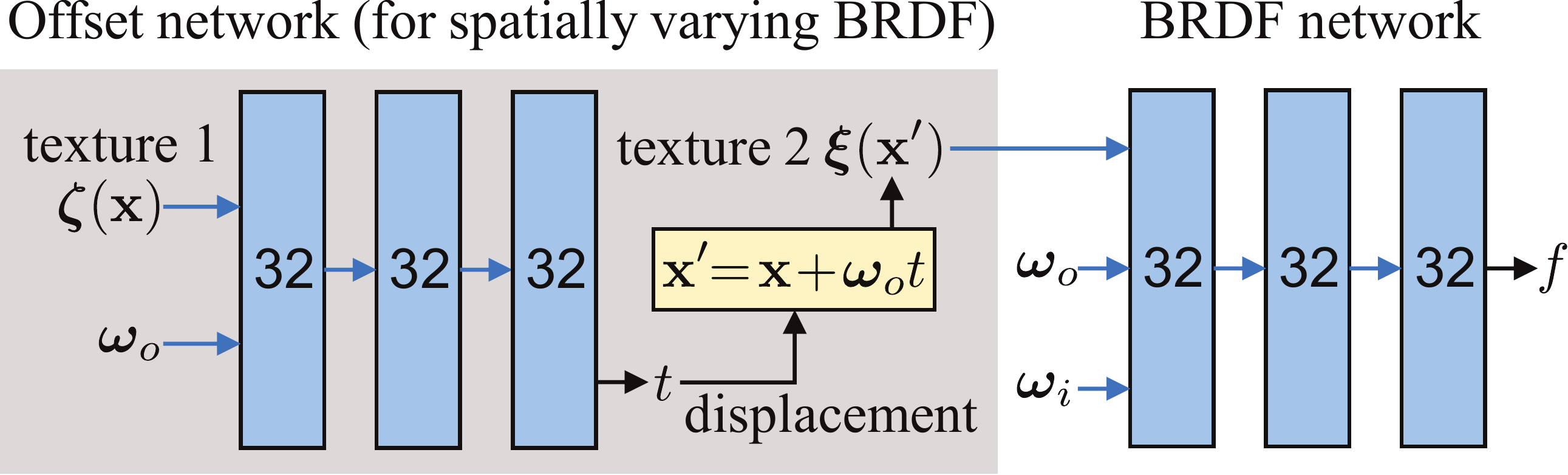}
    \caption{\textbf{Neural BRDF architecture used in the paper},
    which consists of an offset network (left) to model the surface displacement and the network that computes BRDF response (right).
    Spatial signals are encoded in two neural textures ($\bm{\zeta},\bm{\xi}$) of feature size 8.
    The displacement network is only used for spatially varying BRDFs.}
    \label{fig:4-neumip}
\end{figure}
\section{Experiments}
\label{sec:experiments}
We evaluate our model on Monte Carlo rendering with neural materials trained over the RGL dataset~\cite{dupuy2018adaptive} and the spatially varying NeuSample materials~\cite{xu2023neusample}.
The neural BRDF architecture is adopted from \citet{xu2023neusample} (\cref{fig:4-neumip}),
but our method works with other BRDFs as long as $\nabla_{\bm{\omega}_i}f$ is well-defined (for gradient descent optimization of $\mathcal{L}_\text{rep}'$).
Because near-diffuse materials can be easily sampled through cosine-weighted sampling,
we mainly compare renderings with specular materials in terms of
variance in mean squared error (MSE) and the full rendering time
in seconds (sec). The speed-up factors in terms of equal-variance
rendering time relative to the analytic mixture is also reported as a
holistic measure of the efficiency. Since the inference speed in our
render is proportional to the sampling count (spp), we simply take
the averaged speed-up factors over different spps.


\subsection{Baselines}
\label{subsec:baselines}
The baseline methods include cosine-weighted sampling (cosine), normalizing flow (normflow)~\cite{xu2023neusample}, diffusion model~\cite{fu2024importance}, and analytic mixtures of Gaussian and Lambertian lobes~\cite{xu2023neusample} (mixture).
All methods inject $\bm{\omega}_o$ and $\mathbf{x}$ in the same way as in \cref{subsec:architecture}.
For fair comparison, we tune the network size for each baseline to match ours (1.0k parameters for $\mathbf{T}$ + 0.7k parameters for $\hat{p}$).
This is to ensure the performance gain is not from excessive model parameters but the representation itself.
Under this setup, we find the performance of flow-based models can be worse than previously reported by \citeauthor{fu2024importance} and \citeauthor{xu2023neusample} mainly because they over-parameterize their models.
As a sanity check,
we compare our baselines with their original implementations in supplementary.

\begin{figure*}[t]
    \centering
    \setlength\tabcolsep{0.0pt}
    \includegraphics[width=0.98\linewidth]{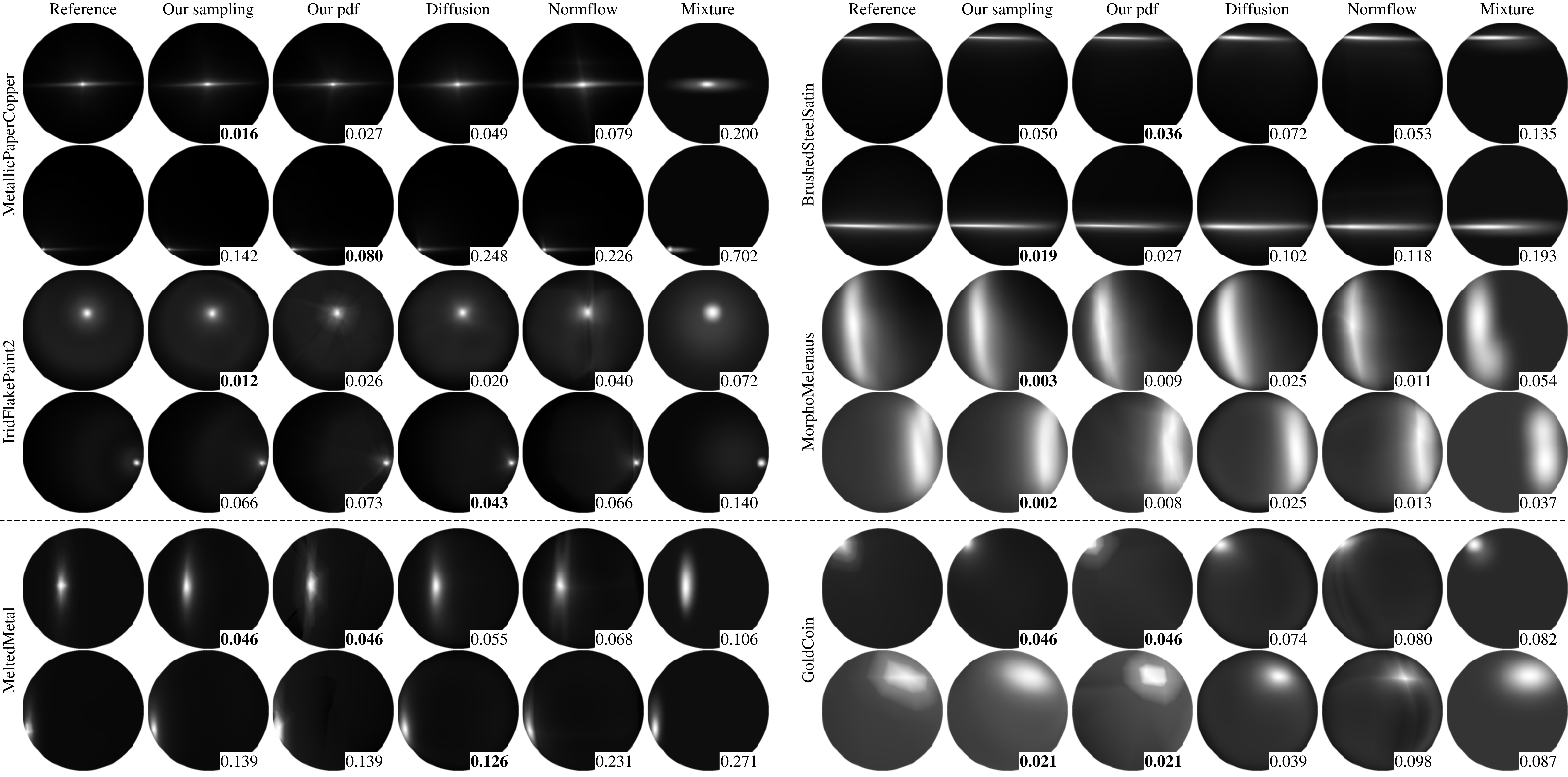}
    \caption{
\textbf{
Visualization of reconstructed pdfs
} 
shows our reparameterization network (our sampling; pdfs generated by binning the samples) most accurately reconstructs the target BRDF distribution (reference),
and the $\hat{p}$ gives reasonable pdf approximation (our pdf).
Every row shows distributions conditioned on different $\mathbf{x},\bm{\omega}_o$ with the numbers denoting the KL-divergence.
Row 1-4 show RGL materials and row 5-6 show NeuSample materials.
    }
    \label{fig:4-lobe}
\end{figure*}
\noindentparagraph{Normalizing flow.}
\citeauthor{xu2023neusample} use 2 coupling MLPs of 1 hidden layer and 32 hidden features to predict rational quadratic splines~\cite{durkan2019neural} of 20 bins.
We change the feature size to 16 and the bin size to 5, giving 1.9k parameters in total.
The base distribution network introduces trainable prior distributions that can be applied to all other methods but shows little improvements on the results.
Therefore, it is dropped for all methods to reduce the computation overhead and for fair comparison.

\noindentparagraph{Diffusion.}
We use the $\mathbb{R}^2$ domain variant of \citeauthor{fu2024importance} and replace their 32-feature MLP by 16 features and 4 hidden layers (1.7k parameters) without the base distribution network.
During rendering, the ODE evaluation takes 4 Euler steps with rectified flow distillation~\cite{liu2022flow} to straighten the ODE path.
Additionally, we show graphs with 80 Euler steps in \cref{fig:4-plot} to demonstrate the best possible variance results, if inference speed is not a criterion.
The pdf visualization in \cref{fig:4-lobe} also uses 80 steps for the same reason.
\noindentparagraph{Analytic mixtures.}
The Gaussian mixture model proposed by \citeauthor{xu2023neusample} is used with an MLP of 4 hidden layers and 16 hidden features (1.7k parameters).

\noindentparagraph{Reference rendering.} 
Cosine-weighted sampling with 4096 spp is used for most materials except for highly specular ones that are sampled through tabulation~\cite{xu2023neusample}.

\subsection{Optimization and inference}
\label{subsec:optimization}
We implement our code using PyTorch~\cite{paszke2019pytorch}, Mitsuba 3~\cite{jakob2022dr}, and CUDA.
The reparameterization network is optimized by $\mathcal{L}_\text{rep}'$ with 0.0005 learning rate,
and the pdf network is optimized by $\mathcal{L}_\text{pdf}$ with 0.001 learning rate.
Both use the Adam optimizer~\cite{kingma2014adam} with 50k steps,
and each step takes 1024 $\mathbf{z}\!\sim\!q$ for each of $1024$ ($\bm{\omega}_o,\mathbf{x}$) samples, resulting in a batch size of $1024^2$.
The training in total takes around 45 minutes for each material on a 3090 GPU with 4GB GPU memory usage.
Pseudocodes of our training and inference is in supplementary. 
All baseline methods are optimized using their original training strategy with $1024^2$ batch size until full convergence.

For inference, the neural BRDF and sampling is implemented as a custom Mitsuba BSDF,
and the rendering uses standard path tracing with MIS.
We write fully-fused CUDA kernels for MLP evaluations as well as forward-mode Jacobian computation,
which noticeably speeds up all methods.
The rendering time is also measured on the 3090 GPU,
and all images are in $800\!\times\!800$ resolution.

\begin{figure*}[p]
    \centering
    \includegraphics[width=0.98\linewidth]{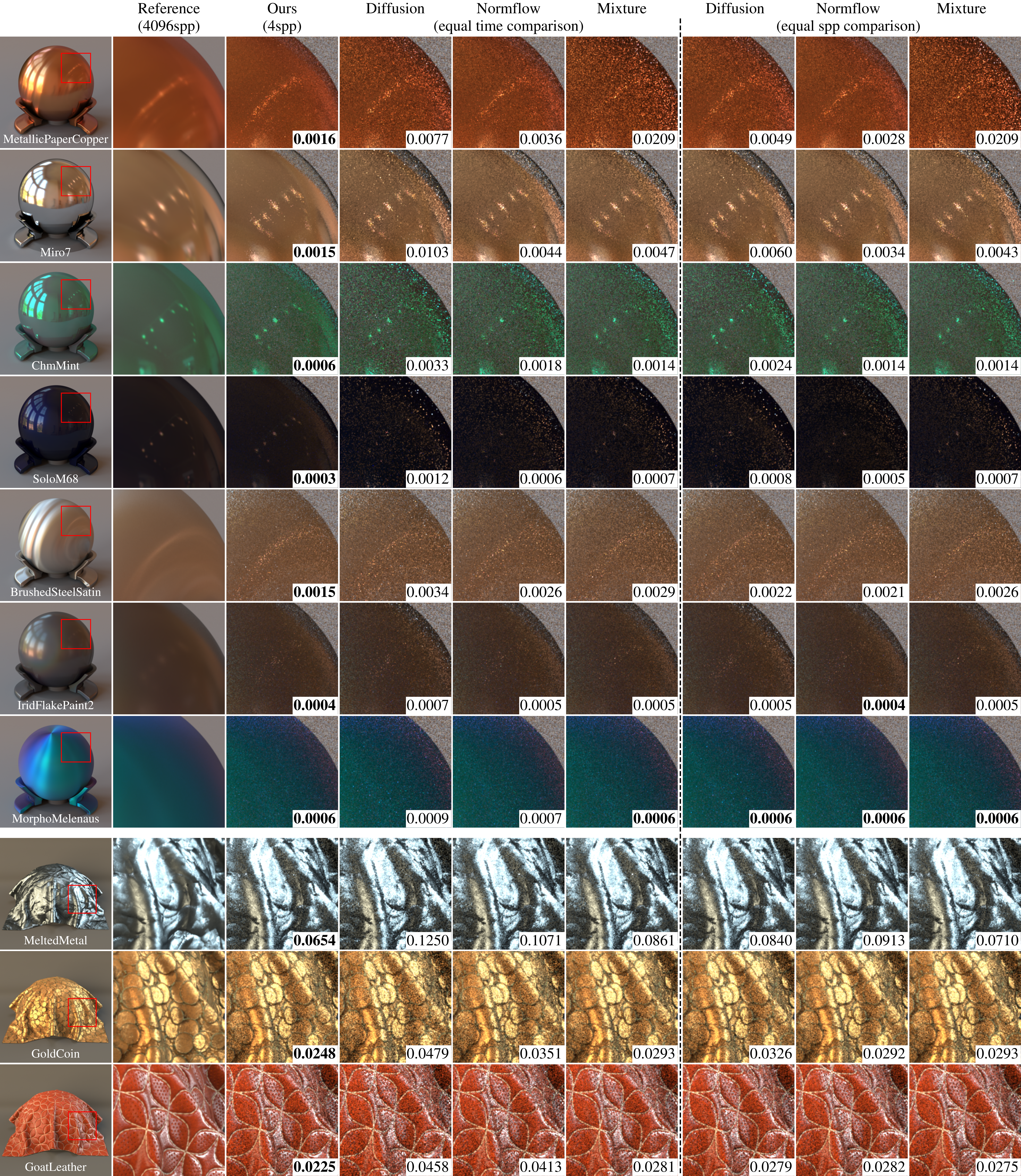}
    \caption{\textbf{Qualitative rendering comparison.}
    The image resolution is $800\!\times\!800$.
    Columns 7-9 are rendered using 4 spp as is ours (column 3).
    Columns 4-6 are rendered with the total sample counts ($4\!\times\!800^2$) adjusted to match our rendering time by a factor of 0.68/0.85/0.98 on RGL materials (rows 1-7) and 0.72/0.85/0.97 on NeuSample materials (rows 8-10; details in supplementary).
    Our renderings demonstrate the least variance in both equal time and equal spp comparison. The numbers show the MSE of the insets.
    }
    \label{fig:4-rendering}
\end{figure*}
\begin{figure*}[p]
    \centering
    \setlength\tabcolsep{1pt}
    \resizebox{0.99\linewidth}{!}{
    \begin{tabular}{cc}
        \multicolumn{2}{c}{\qquad\textbf{RGL materials}}\\
         \includegraphics[width=0.55\linewidth]{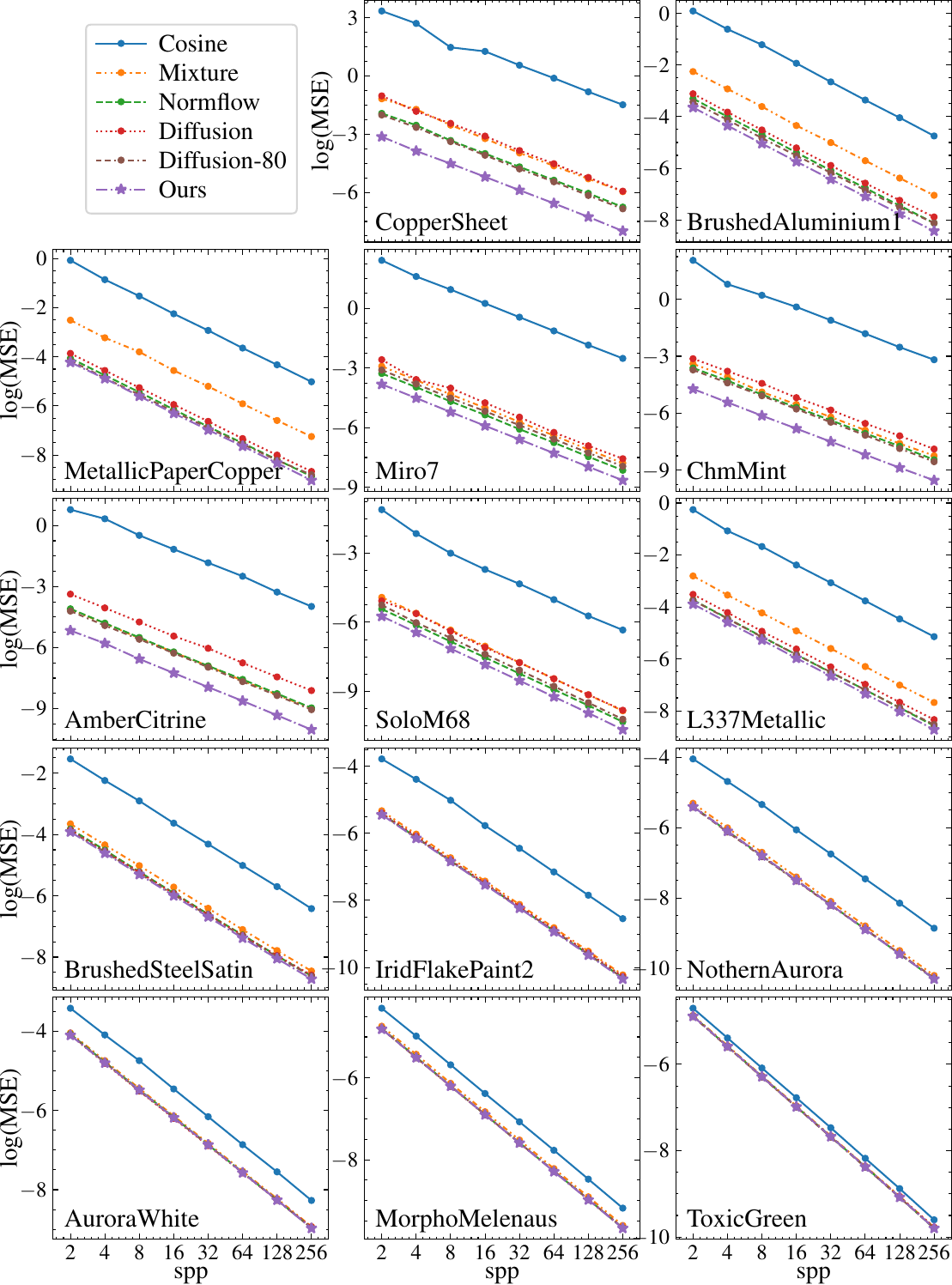}&
         \includegraphics[width=0.55\linewidth]{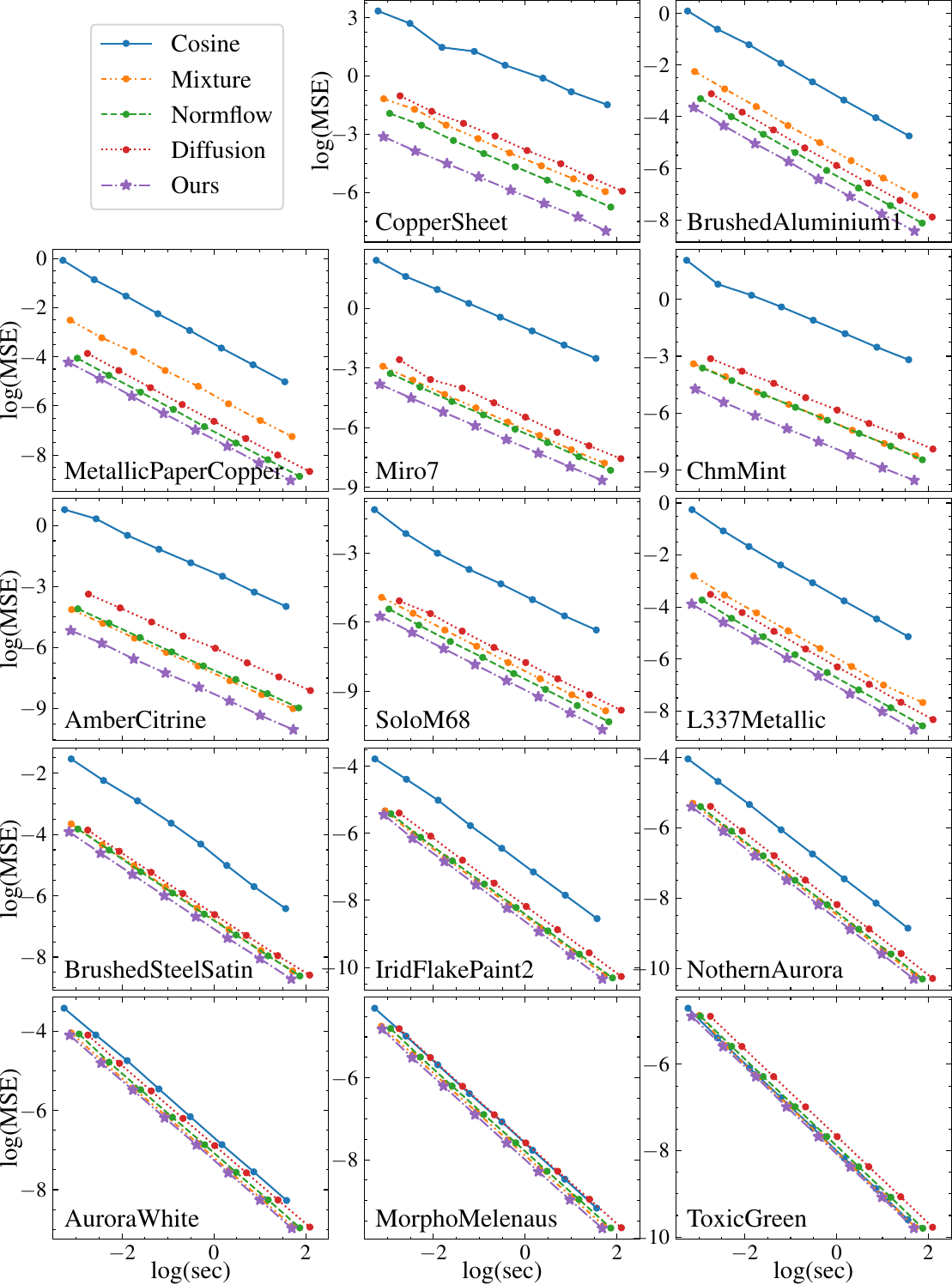}\\
         \qquad log spp vs. log MSE & \qquad log MSE vs. log time (sec)\\
         [12pt]
         \multicolumn{2}{c}{\qquad\textbf{NeuSample materials}}\\
         \includegraphics[width=0.55\linewidth]{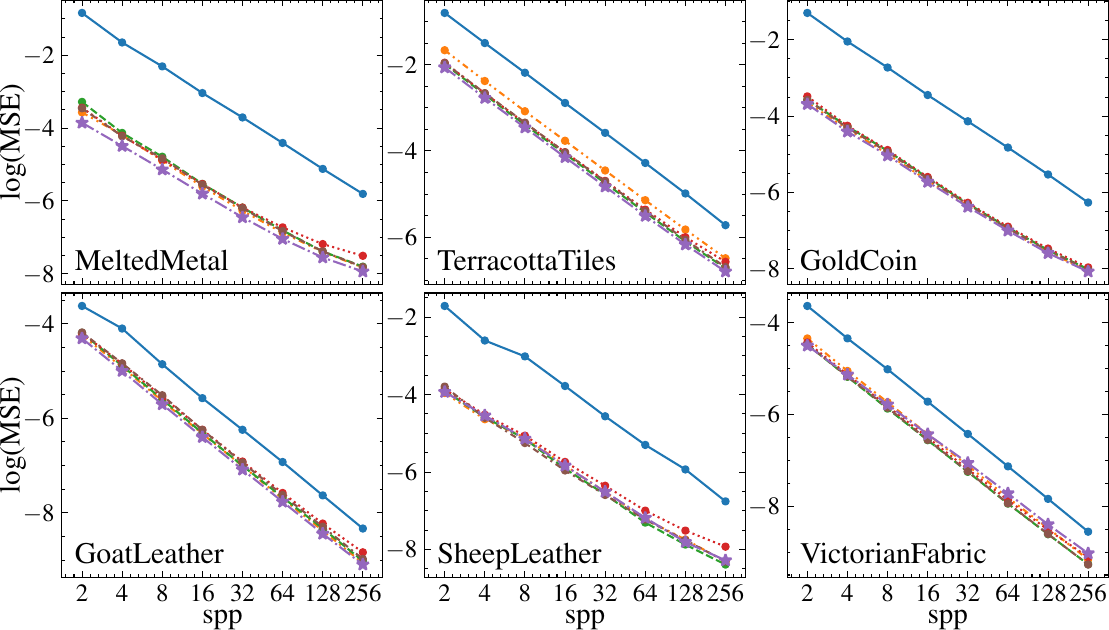}&
         \includegraphics[width=0.55\linewidth]{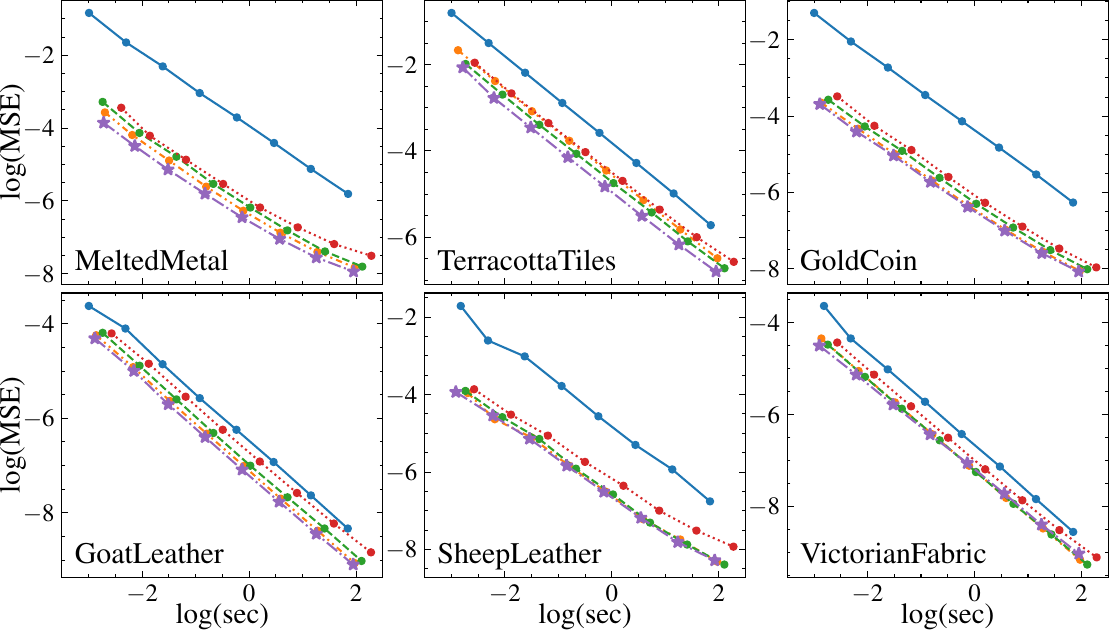}\\
         \qquad log spp vs. log MSE & \qquad log MSE vs. log time (sec)\\
    \end{tabular}
    }
    \caption{\textbf{Convergence graphs.}
    Columns 1-3 show the log-log plot of the rendering MSE with respect to the spp.
    Columns 4-6 show the log-log plot of the rendering MSE with respect to the rendering time.
    Our method achieves the overall best variance reduction in both comparisons, 
    and the improvements are more significant on specular materials (rows 1-3) than the rest.
    "Diffusion" is evaluated using 4 ODE steps, and "diffusion-80" takes 80 ODE steps.
    }
    \label{fig:4-plot}
\end{figure*}
\subsection{Results}
\label{subsec:results}
\noindentparagraph{Accuracy of reconstructed distributions.}
Except for analytic mixtures whose shapes are restricted to Gaussian blobs, 
both flow-based models and our sampling approximately recover the reference distributions (\cref{fig:4-lobe}).
However, on closer inspection, we see that the normalizing flow underfits \textit{IridFlakePaint2} and \textit{MetalicPaperCopper} owing to the limited expressivity of the spline functions it uses.
Our sampling network can learn nearly arbitrary one-to-one probability transforms so demonstrates uniformly lower KL-divergence (numbers in \cref{fig:4-lobe}) even though our network size is smaller.
While the diffusion model in most case outperforms the normalizing flow when using a very large number of parameters~\cite{fu2024importance},
it produces similar results as the normalizing flow when using a similar network. 
$\hat{p}$ (our pdf) does not need to be precise for calculating the MIS weights,
while it sometimes still gives better KL-divergence than the baselines.

\noindentparagraph{Rendering variance analysis.}
As shown in the equal-spp renderings in \cref{fig:4-rendering},
our model noticeably reduces the noise on specular materials like \textit{MetallicPaperCopper} and \textit{ChmMint},
while the improvements on rough materials (\eg \textit{MorphoMelenaus}) are less significant.
This is because their BRDFs are smooth and flat and therefore do not exhibit too much variance in renderings.
The above observation is quantitatively visualized in the equal-spp convergence graphs (\cref{fig:4-plot} left),
showing our variance reduction is either better or comparable with the baselines.
Note that using 4 Euler steps for the diffusion ODE evaluation (diffusion) may introduce noticeable discretization error, leading to even worse results than the analytic mixture (\eg \textit{AmberCitrine}).
However, the 80-step diffusion model (diffusion-80 in \cref{fig:4-plot}) is far slower than the rest of the methods so will not be further compared in the efficiency test below,
and in any case, the diffusion model is still worse than our method.
In supplementary, we also provide convergence graphs on renderings with constant light similar to \citeauthor{xu2023neusample} and a discussion of grazing angle behaviors similar to \citeauthor{fu2024importance}.

\begin{table}[t]
    \centering
    \caption{
    \textbf{Timing of different rendering stages.}
    Our BRDF sampling (including $\hat{p}(\bm{\omega}_i)$ evaluation of the sample) is comparable with analytic mixture in speed,
    and our pdf evaluation (for emitter samples) is even faster.
    Other parts of the rendering (other) take similar time.
    The time is recorded for 4 spp renderings averaged over all materials with the best mark in \textbf{bold}.
    }
    \setlength\tabcolsep{2.5pt}
\resizebox{0.99\linewidth}{!}{
\begin{tabular}{c l cccc}
\toprule
\multirow{2}{*}{\textbf{Materials}}& \multirow{2}{*}{\textbf{Rendering stage}} & \textbf{Mixture} & \textbf{Normflow} & \textbf{Diffusion} & \textbf{Ours}\\
& &\multicolumn{4}{c}{\textbf{sec$\downarrow$}}\\
\midrule
\multirow{4}{*}{RGL} &
BRDF sampling&
\textbf{0.0063} & 0.0128 & 0.0288 & 0.0065
\\
&Pdf evaluation&
0.0031 & 0.0098 & 0.0181 & \textbf{0.0021}
\\
&Other&
0.0774 & 0.0776 & 0.0781 & \textbf{0.0761}
\\
&Total&
0.0868 & 0.1002 & 0.1250 & \textbf{0.0847}
\\
\midrule
\multirow{4}{*}{NeuSample} &
BRDF sampling&
\textbf{0.0082} & 0.0157 & 0.0307 & 0.0083
\\
&Pdf evaluation&
0.0044 & 0.0116 & 0.0198 & \textbf{0.0030}
\\
&Other&
0.1000 & 0.1014 & 0.1025 & \textbf{0.0984}
\\
&Total&
0.1126 & 0.1288 & 0.1530 & \textbf{0.1097}
\\
\bottomrule
\end{tabular}
}
    \label{tab:4-timing}
\end{table}
\begin{table}[t]
    \centering
        \caption{
    \textbf{Speed up relative to "Mixture"} shows our method achieves the best performance in terms of equal-variance rendering time.
    }
    \setlength\tabcolsep{2.pt}
    \resizebox{0.99\linewidth}{!}{
    \begin{tabular}{p{6pt}l ccccc}
    \toprule
&& \textbf{Cosine} & \textbf{Mixture} & \textbf{Normflow} & \textbf{Diffusion} & \textbf{Ours}\\
\midrule
&\textbf{RGL materials} 
&\multicolumn{5}{c}{\textbf{Speed up relative to Mixture$\uparrow$}}\\
\midrule
\multirow{8}{*}{\rotatebox[origin=c]{90}{More specular}}
&CopperSheet & 0.01 & 1.00 & 1.89 & 0.64 & \textbf{7.14} \\
&BrushedAluminium1 & 0.11 & 1.00 & 2.49 & 1.66 & \textbf{4.16} \\
&MetallicPaperCopper & 0.12 & 1.00 & 4.23 & 2.82 & \textbf{5.96} \\
&Miro7 & 0.01 & 1.00 & 1.22 & 0.56 & \textbf{2.50} \\
&ChmMint & 0.01 & 1.00 & 1.03 & 0.48 & \textbf{3.80} \\
&AmberCitrine & 0.01 & 1.00 & 0.85 & 0.30 & \textbf{2.78} \\
&SoloM68 & 0.04 & 1.00 & 1.45 & 0.70 & \textbf{2.32} \\
&L337Metallic & 0.09 & 1.00 & 2.17 & 1.39 & \textbf{3.01} \\
\midrule
\multirow{6}{*}{\rotatebox[origin=c]{90}{Less specular}}
&BrushedSteelSatin & 0.13 & 1.00 & 1.03 & 0.83 & \textbf{1.36} \\
&IridFlakePaint2 & 0.22 & 1.00 & 0.93 & 0.73 & \textbf{1.16} \\
&NothernAurora & 0.30 & 1.00 & 0.94 & 0.74 & \textbf{1.13} \\
&AuroraWhite & 0.59 & 1.00 & 0.89 & 0.72 & \textbf{1.07} \\
&MorphoMelenaus & 0.75 & 1.00 & 0.91 & 0.72 & \textbf{1.11} \\
&ToxicGreen & 0.98 & 1.00 & 0.88 & 0.70 & \textbf{1.06} \\
\midrule
&\textbf{NeuSample materials}
&\multicolumn{5}{c}{\textbf{Speed up relative to Mixture$\uparrow$}}\\
\midrule
\multirow{6}{*}{\rotatebox[origin=c]{90}{Less specular}}
&MeltedMetal & 0.10 & 1.00 & 0.82 & 0.65 & \textbf{1.23} \\
&TerracottaTiles & 0.48 & 1.00 & 1.16 & 0.92 & \textbf{1.47} \\
&GoldCoin & 0.14 & 1.00 & 0.82 & 0.67 & \textbf{1.06} \\
&GoatLeather & 0.53 & 1.00 & 0.86 & 0.66 & \textbf{1.09} \\
&SheepLeather & 0.17 & 1.00 & 0.93 & 0.62 & \textbf{1.03} \\
&VictorianFabric & 0.58 & \textbf{1.00} & 0.99 & 0.78 & \textbf{1.00}\\
    \bottomrule
    \end{tabular}
    }
    \label{tab:4-efficiency}
\end{table}
\noindentparagraph{Inference speed comparison.}
It can be seen in \cref{tab:4-timing} that our BRDF sampling is around $4\times$ faster than the diffusion model as the latter needs 4 network evaluations (ODE steps),
and it is $2\times$ faster than the normalizing flow whose spline evaluation creates control divergence.
In contrast, computing $\mathbf{T},\,\mathbf{J_T},\,\hat{p}$ in our model requires only tensor core operations that maximizes parallelism.
This is also likely why our sampling is only slightly slower than the analytic mixture,
but our pdf evaluation is faster owing to the small network size in $\hat{p}$,
resulting in the overall lowest inference time.

\noindentparagraph{Overall performance.}
The fast inference speed allows our method to allocate more samples in the given time to further reduce the rendering variance (\cref{fig:4-plot} right).
In contrast, flow-based models are less efficient and produce even more noise in the equal-time rendering comparison (\cref{fig:4-rendering}).
Quantitatively, our model is 2.32-7.14$\times$ more efficient than the analytic mixture on the first 8 materials in \cref{tab:4-efficiency},
giving 1.39-3.78$\times$ speedup of the second best methods;
and it is still slightly better on the other materials (ranging from 1$\times$ to 1.47$\times$).
Note that the analytic mixture can perform better than the flow-based models on many BRDFs because of its fast speed.
While previous works suggest the flow-based models are always better~\cite{xu2023neusample,fu2024importance},
they record the inference in Pytorch with larger networks rather than the optimized CUDA kernels we use with similar model size.

\subsection{Ablation study}
\label{subsec:ablation}

\begin{figure}[t]
    \centering
    \includegraphics[width=0.75\linewidth]{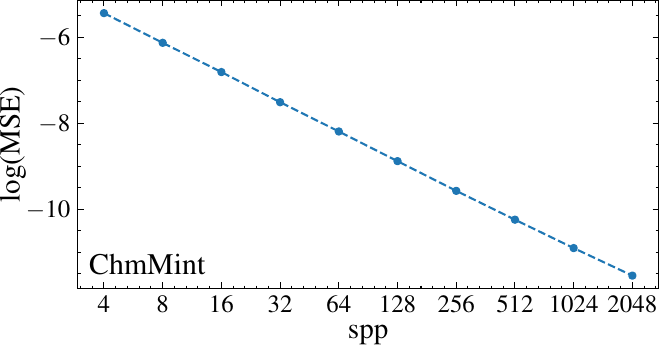}
    \caption{\textbf{Convergence graph extending to a longer rendering period for our method} shows a linear error deduction in log-log space, which indicates our estimator is unbiased.}
    \label{fig:6-unbias}
\end{figure}
\noindentparagraph{Empirical test on unbiasedness.}
Figure \ref{fig:6-unbias} shows a longer MSE-spp log-log plot till 2048 spp of our method on \textit{ChmMint}, which is close to a straight line that suggests the unbiasedness of our method is usually guaranteed.

\begin{table}[t]
    \centering
        \caption{
    \textbf{Quantitative comparison with the tabulated importance sampling on \textit{Miro7} material} suggests our method produces comparable variance reduction and inference speed.
    }
    \setlength\tabcolsep{4.5pt}
    \resizebox{0.9\linewidth}{!}{
    \begin{tabular}{l ccccccc}
    \toprule
    \textbf{spp} & 4 & 8 & 16 & 32 & 64 & 128 & 256 \\
    \midrule
    &\multicolumn{7}{c}{\textbf{log(MSE)$\downarrow$}}\\
    \midrule
    \textbf{Tabulated} & -4.49 & -5.18 & \textbf{-5.90} & \textbf{-6.59} & \textbf{-7.30} & \textbf{-8.01} & \textbf{-8.70}\\
    \textbf{Ours} & \textbf{-4.51} & \textbf{-5.21} & \textbf{-5.90} & \textbf{-6.59} & -7.28 & -7.97 & -8.65\\
    \midrule
    &\multicolumn{7}{c}{\textbf{sec$\downarrow$}}\\
    \midrule
    \textbf{Tabulated} & \textbf{0.04} & \textbf{0.08} & \textbf{0.17} & \textbf{0.34} & \textbf{0.66} & \textbf{1.33} & \textbf{2.65}\\
    \textbf{Ours} & 0.05 & 0.09 & 0.18 & 0.36 & 0.71 & 1.42 & 2.87\\
    \bottomrule
    \end{tabular}
    }
    \label{tab:6-tabulation}
\end{table}
\noindentparagraph{Comparison with tabulated importance sampling.}
We conduct variance-speed comparison with Mitsuba's~\cite{jakob2022dr} tabulated importance sampling on \textit{Miro7}.
It can be seen in \cref{tab:6-tabulation} that our method has very similar variance reduction effect as the tabulated importance sampling that gives a near-optimal discretization of the ground truth pdf.
The tabulated sampling is faster likely because it is native to the Mitsuba pipeline,
but the difference is very small as fetching the tabulated texture can be as inefficient as the network execution~\cite{fu2024importance}.
Meanwhile, our storage cost is much smaller (1.7KB vs. 4.4MB).

\noindentparagraph{Fully-fused CUDA kernel speed-up.}
Table~\ref{tab:4-cuda} shows the rendering time with native Pytorch and our fully-fused CUDA kernels.
Unlike in \cref{tab:4-timing},
the BRDF sampling in Pytorch dominates the rendering time for flow-based models.
Our method in Pytorch is slightly slower than the mixture model because its Jacobian evaluation is inefficient,
and the situation is worse for the diffusion model that computes more Jacobians.
Therefore, it is necessary to use the fully-fused CUDA kernels to maintain small BRDF sampling overhead.

\begin{figure}[t]
    \centering
    \includegraphics[width=0.99\linewidth]{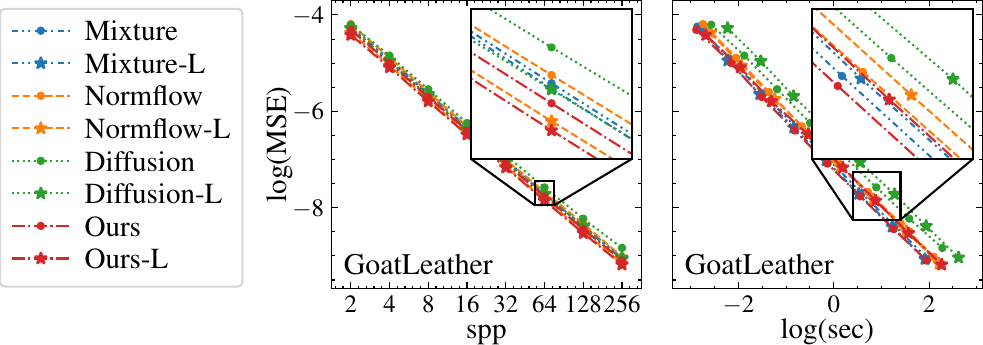}
    \caption{\textbf{Ablation on network architecture}
    shows increasing network size ("-L") leads to better equal-spp results (middle), and our model (Ours-L) still gives the best variance reduction. The larger models are slower so have decreased efficiency (right plot).
    }
    \label{fig:4-variant}
\end{figure}

\noindentparagraph{Network architecture.}
Figure~\ref{fig:4-variant} shows the performance trade-off of each model when using 32 hidden features (and also 20 bins for normflow) on the \textit{GoatLeather} materials.
With larger MLPs, these models (marked with “-L”) converge to lower equal-spp variance (\cref{fig:4-variant} middle), while the improvement on the analytic mixture is subtle.
On the other hand, the hidden features no longer fit into the CUDA tensor core,
so they need to be evicted into GPU shared memory more often during MLP evaluations,
This slows down the inference that leads to worse equal-time performance (\cref{fig:4-variant} right).

\noindentparagraph{Limitations.}
Our optimization relies on gradients of the target distribution so suffers from the mode-seeking issue if the distribution has strong disconnections (\cref{fig:5-limitations}).
Meanwhile, the Jacobian computation is impractical for high dimensional distributions, which may be solved by autoregressive modeling~\cite{van2016conditional}.
BRDFs are in low dimension, and multi-modal BRDFs (\eg fibers) are always smooth.
Therefore, both issues are not relevant to the BRDF sampling application.
\begin{table}[t]
    \centering
        \caption{
    \textbf{Ablation on different inference backends.}
    The fully-fused CUDA kernels noticeably speed up all the methods.
    }
    \setlength\tabcolsep{2.5pt}
\resizebox{0.99\linewidth}{!}{
\begin{tabular}{c l cccc}
\toprule
\multirow{2}{*}{\textbf{Material}}& \multirow{2}{*}{\textbf{Backend}} & \textbf{Mixture} & \textbf{Normflow} & \textbf{Diffusion} & \textbf{Ours}\\
& &\multicolumn{4}{c}{\textbf{sec$\downarrow$}}\\
\midrule
\multirow{2}{*}{RGL} &
Fully-fused CUDA&
0.0868 & 0.1002 & 0.1250 & \textbf{0.0847}
\\&
Native Pytorch &
\textbf{0.1128} & 0.4568 & 0.5480 & 0.1223
\\
\midrule
\multirow{2}{*}{NeuSample} &
Fully-fused CUDA&
0.1126 & 0.1288 & 0.1530 & \textbf{0.1097}
\\&
Native Pytorch &
\textbf{0.1567} & 0.4848 & 0.6808 & 0.1691
\\
\bottomrule
\end{tabular}
}
    \label{tab:4-cuda}
\end{table}
\begin{figure}[t]
    \centering
    \setlength\tabcolsep{0.4pt}
    \resizebox{0.99\linewidth}{!}{
    \begin{tabular}{cccc}
         \includegraphics[width=0.28\linewidth]{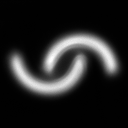}&
         \includegraphics[width=0.28\linewidth]{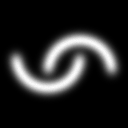}&
         \includegraphics[width=0.28\linewidth]{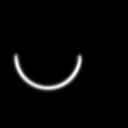}&
         \includegraphics[width=0.28\linewidth]{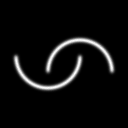}\\
         Ours & Reference & Ours & Reference
    \end{tabular}
    }
    \caption{\textbf{Mode-seeking issue with our model.}
    Multi-modal distributions can be correctly constructed if there are gradients that allow $\mathcal{L}_\text{rep}'$ to propagate samples (column 1-2),
    but it fails when two modes are completely disconnected (column 4),
    where our sampling is stuck in a single mode (column 3).
    }
    \label{fig:5-limitations}
\end{figure}
\section{Conclusion and Future Work}
\label{sec:conclusion}
We have demonstrated a novel BRDF importance sampling strategy by learning an optimal reparameterization of the BRDF integral.
This approach allows the use of arbitrary $C^1$-continuous MLPs to sample BRDFs in a single step, enabling more efficient and effective variance reduction in Monte Carlo rendering.
Beyond rendering, our model has the potential to benefit other low-dimensional applications that require importance sampling.

\begin{acks}
This work was supported in part by NSF grant 2212085 and the Ronald L. Graham Chair.
Ramamoorthi acknowledges a part-time appointment at NVIDIA.
We also acknowledge gifts from Adobe, Google, Qualcomm and Rembrand and the UC San Diego Center for Visual Computing.
\end{acks}

\bibliographystyle{ACM-Reference-Format}
\bibliography{bibliography}
\clearpage
\newpage
\begin{appendix}
    \begin{algorithm}[t!]
\SetAlgoNoLine
\For{each training step}{
$\bm{\omega}_o,\mathbf{x}\sim$ sample condition\\
$\mathbf{z}\sim q$\\
$\bm{\omega}_i = \mathbf{T}(\mathbf{z}|\bm{\omega}_o,\mathbf{x}),\;\mathbf{J}=\text{Jacobian}(\bm{\omega}_i,\mathbf{z})$\\
$p=f(\mathbf{x},\bm{\omega}_i,\bm{\omega}_o)\max(\det\mathbf{J},0)$\\
$\bm{\omega}_i' = \mathbf{I}(\mathbf{z}),\;\mathbf{J}'=\text{Jacobian}(\bm{\omega}_i',\mathbf{z})$\\
$p'=f(\mathbf{x},\bm{\omega}_i',\bm{\omega}_o)\left|\det\mathbf{J}'\right|$\\
$loss=\text{mean}\left(-\log(p(1-\alpha)+p'\alpha)\right)$\\
Take gradient descent step on $\nabla_\mathbf{T}loss$ 
}
\caption{Training $\mathbf{T}$}
\label{alg:6-trainingT}
\end{algorithm}

\begin{algorithm}[t]
\SetAlgoNoLine
\For{each training step}{
$\bm{\omega}_o,\mathbf{x}\sim$ sample condition\\
$\mathbf{z}\sim q$\\
$\bm{\omega}_i = \mathbf{T}(\mathbf{z}|\bm{\omega}_o,\mathbf{x}),\;\mathbf{J}=\text{Jacobian}(\bm{\omega}_i,\mathbf{z})$\\
$loss$=$\text{mean}\left(\left(\hat{p}(\bm{\omega}_i|\bm{\omega}_o,\mathbf{x})\left|\det\mathbf{J}\right|-q(\mathbf{z})\right)^2\right)$\\
Take gradient descent step on $\nabla_\mathbf{\hat{p}}loss$ 
}
\caption{Training $\hat{p}$}
\label{alg:6-trainingp}
\end{algorithm}%
\begin{algorithm}[t]
\SetAlgoNoLine
\KwIn{$\bm{\omega}_o$, $\mathbf{x}$, $w$, $L_o$, path tracing throughput $\beta$}
\KwOut{next direction $\bm{\omega}_i$, updated $w,\beta,L_o$}
$L_e=$ get surface emission\\
$L_o = L_o + w L_e \beta$\\
$L'_e,\bm{\omega}'_i=$ sample emitter\\
$w_e=\hat{p}^2_e(\bm{\omega}'_i|\bm{\omega}_o,\mathbf{x})/(p^2(\bm{\omega}'_i|\bm{\omega}_o,\mathbf{x})+p_e^2(\bm{\omega}'_i|\bm{\omega}_o,\mathbf{x}))$\\
$\beta' = \beta \times f(\mathbf{x},\bm{\omega}'_i,\bm{\omega}_o)/p_e(\bm{\omega}'_i|\bm{\omega}_o,\mathbf{x})$\\
$L_o = L_o + w_e L'_e \beta'$\\
$\mathbf{z}\sim q$\\
$\bm{\omega}_i=\mathbf{T}(\mathbf{z}|\bm{\omega}_o,\mathbf{x}),\;\mathbf{J}=\text{Jacobian}(\bm{\omega}_i,\mathbf{z})$\\
$\beta=\beta \times f(\mathbf{x},\bm{\omega}_i,\bm{\omega}_o)\left|\det\mathbf{J}\right|/q(\mathbf{z})$\\
$w=\hat{p}^2(\bm{\omega}_i|\bm{\omega}_o,\mathbf{x})/(\hat{p}^2(\bm{\omega}_i|\bm{\omega}_o,\mathbf{x})+p_e^2(\bm{\omega}_i|\bm{\omega}_o,\mathbf{x}))$\\
\Return
\caption{Inference}
\label{alg:6-inference}
\end{algorithm}%
\section{Additional Derivations}
\subsection{$\mathcal{L}_\text{nll}=$ reverse KL-Divergence at convergence}
If the inverse $\mathbf{T}^{-1}$ exists,
$\mathcal{L}_\text{nll}$ produces the same gradient as the reverse KL-divergence $\int\!p\log\left(\frac{p}{f/F}\right)\!d\mathbf{z}$:
\begin{align}
\nabla_\mathbf{T}\mathcal{L}_\text{nll}=
\nabla_\mathbf{T}\mathcal{L}_\text{nll}
+\nabla_\mathbf{T}\!\!\int\!\!q(\mathbf{z})
\log\left(Fq(\mathbf{z})\right)\mathrm{d}\mathbf{z}
\tag*{}
\\
=\nabla_\mathbf{T}\!\!\int\!\!q(\mathbf{z})\log\left(
\frac{Fq(\mathbf{z})}
{f(\mathbf{T}(\mathbf{z}))|\det\mathbf{J}_\mathbf{T}(\mathbf{z})|}
\right)\!\mathrm{d}\mathbf{z}
\tag*{}
\\
=\nabla_\mathbf{T}\!\!\int\!\!
q(\mathbf{T}^{-1})\left|\det\mathbf{J}_{\mathbf{T}^{-1}}\right|
\log\left(
\frac{|\!\det\mathbf{J}_{\mathbf{T}^{-1}}\!|q(\mathbf{T}^{-1})}
{f/F}
\right)\!\mathrm{d}\bm{\omega}_i
\tag*{}
\\
=\nabla_\mathbf{T}\!\!\int\!\!p\log\left(\frac{p}{f/F}\right)\mathrm{d}\mathbf{z}.
\tag*{}
\end{align}%
Therefore, the two losses are equivalent when $\mathbf{T}$ has a valid inverse that corresponds to the end of the optimization.

\subsection{$\mathcal{L}_\text{rep}'$ is an upper bound of $\mathcal{L}_\text{rep}$}
It can be seen that:
\begin{align}
\left|\det\mathbf{J}_\mathbf{T}\right|\geq\max(\det\mathbf{J}_\mathbf{T},0)\tag*{}\\
\Rightarrow
(1-\alpha)f(\mathbf{T})\left|\det\mathbf{J}_\mathbf{T}\right|+\alpha f(\mathbf{I})\left|\det\mathbf{J}_\mathbf{T}\right|\tag*{}\\
\geq
(1-\alpha)f(\mathbf{T})\max(\det\mathbf{J}_\mathbf{T},0)+\alpha
f(\mathbf{I})\left|\det\mathbf{J}_\mathbf{T}\right|.\tag*{}
\end{align}%
Since $-\log(\cdot)$ is a monotonically decreasing function, we have $\mathcal{L}'_\text{rep}\geq\mathcal{L}_\text{rep}$.
\begin{figure}[t]
    \centering
    \setlength\tabcolsep{0.4pt}
    \resizebox{0.99\linewidth}{!}{
    \begin{tabular}{cccc}
         \includegraphics[width=0.28\linewidth]{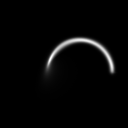}&
         \includegraphics[width=0.28\linewidth]{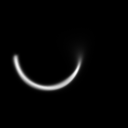}&
         \includegraphics[width=0.28\linewidth]{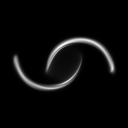}&
         \includegraphics[width=0.28\linewidth]{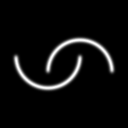}\\
        w/ identity map & w/ gradient clip & w/ gradient clip & \multirow{2}{*}{Reference}\\[-1.2pt]
        initialization & ($10^{-3}$ max norm) & ($10^{-4}$ max norm)
    \end{tabular}
    }
    \caption{\textbf{Mode-seeking can be avoided} by initializing the network to be an identity map then applying gradient clipping during training (3rd image).
    The gradient clipping max norm needs to be very small (2nd vs. 3rd image), which may slow down the convergence.
    }
    \label{fig:6-mode}
\end{figure}%
\section{Resolve Mode-seeking}
Let $\mathbf{T}(\mathbf{z},\bm{\theta})$ denote the reparameterization network with input $\mathbf{z}$ and network parameters $\bm{\theta}$.
We prevent mode-seeking by first initializing $\mathbf{T}$ to be an identity map to cover the target distribution,
which ensures all modes are taken into account during optimization.
A noisy gradient descent step can still drastically perturb the $\mathbf{T}$ to be stuck in a local mode (1st image of \cref{fig:6-mode}).
We therefore further constrain $\mathbf{T}(\mathbf{z},\bm{\theta}_{t+1}), \mathbf{T}(\mathbf{z},\bm{\theta}_{t})$ to be close to each other for network parameters in two adjacent training steps $\bm{\theta}_{t+1}, \bm{\theta}_t$.
By mean value theorem, this essentially requires $\bm{\theta}_{t+1}-\bm{\theta}_t$ to be small:
\begin{equation}
    \mathbf{T}(\mathbf{z},\bm{\theta}_{t+1})=\mathbf{T}(\mathbf{z},\bm{\theta}_t)+\nabla_\theta\mathbf{T}(\mathbf{z},\bm{\theta}'_t)(\bm{\theta}_{t+1}-\bm{\theta}_t),
    \; \bm{\theta}'_t\in[\bm{\theta}_{t},\bm{\theta}_{t+1}],
\end{equation}
which can be achieved by gradient clipping.
As demonstrated in the second image of \cref{fig:6-mode}, our model successfully learns to sample the strongly-disconnected distribution using the two strategies above.
Tuning the hyperparameters for gradient clipping, however, is very tricky (3rd image of \cref{fig:6-mode}).

\section{Implementation details}
\cref{alg:6-trainingT,alg:6-trainingp} show the pseudocode for training $\mathbf{T}$ and $\hat{p}$.
\cref{alg:6-inference} shows the pseudocode of our inference in a path-tracing renderer.

\paragraph{Baseline training.}
The analytic mixture and the normalizing flow are trained by optimizing forward KL-divergence with ground truth samples $\bm{\omega}_i$ generated by the online tabulation~\cite{xu2023neusample}.
The diffusion model is trained over the IADB objective~\cite{heitz2023iterative} that is equivalent to the flow matching loss~\cite{lipman2022flow}.
While \citet{fu2024importance} use MCMC to generate the ground truth $\bm{\omega}_i$, it can only give a finite dataset that is insufficient for training spatially varying BRDFs.
Instead, we use the online tabulation method that in theory can create infinite combinations of $\bm{\omega}_i$ for training.

\begin{figure}[t]
    \centering
    \begin{tabular}{c}
        \includegraphics[width=0.8\linewidth]{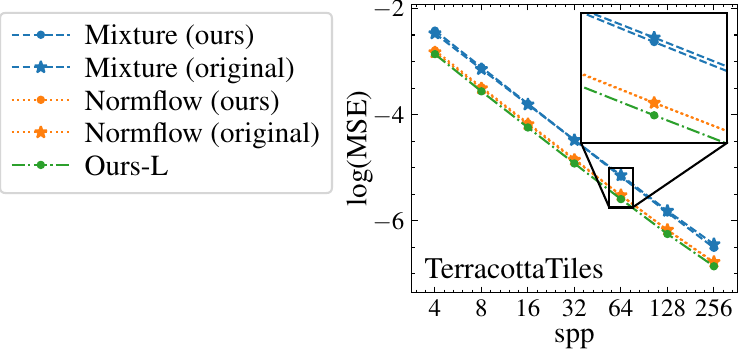}\\
        \includegraphics[width=0.8\linewidth]{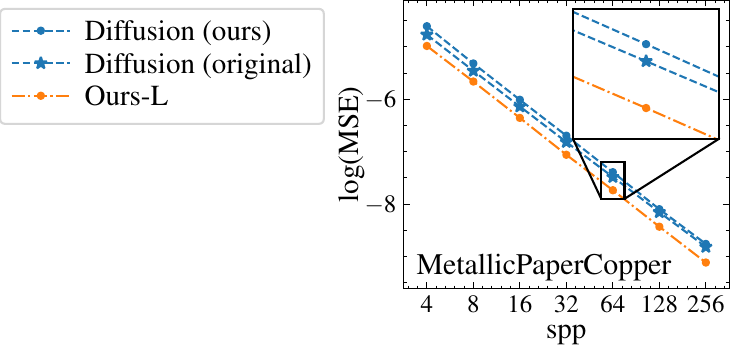}
    \end{tabular}
    \caption{\textbf{Our baselines vs. original baselines as equal-spp convergence graphs,}
    showing our implementation is consistent with the original implementation on all baselines (mixture, normflow, and diffusion).
    Here, we adjust the feature size of our reparameterization model (Ours-L) to 32 to match the network size of the baseline models.
    }
    \label{fig:6-sanity-1}
\end{figure}
\paragraph{Baseline implementation sanity check.}
We compare our trained baselines with the baselines using their original papers' training weights~\cite{xu2023neusample,fu2024importance} on the \textit{TerracottaTiles} material and the \textit{MetallicPaperCopper} material.
It can be seen in \cref{fig:6-sanity-1} that our baselines produce similar MSE-spp log-log plots as the original models, suggesting the correctness of our implementation.
Meanwhile, our reparameterization model (Ours-L) demonstrates the lowest variance under similar network size, which is also consistent with the observation in Sec.~4.3 of the paper.
Note that the diffusion models of \citeauthor{fu2024importance} are trained on the raw RGL materials~\cite{dupuy2018adaptive} rather than their neural BRDFs,
but the difference between the two representations is negligible.

\paragraph{Equal time renderings.}
Taking the diffusion baseline on the RGL materials as an example that has an average of $0.68\!\times\!4\!=\!2.72$ spp,
we first allocate 2 spp for each pixel then distribute the remaining $\lfloor0.72\!\times\!800^2\rfloor$ samples to random pixels without replacement.

\section{Additional results}
Figure~\ref{fig:6-plot} shows additional visualizations of reconstructed pdfs for each method. 
Figures~\ref{fig:6-rendering1} and~\ref{fig:6-rendering2} show our full image renderings with 8 spp on materials evaluated in the paper.
While more specular materials are available in the RGL dataset~\cite{dupuy2018adaptive}, 
they differ mainly in color (e.g. \textit{ChmnLightBlue} vs. \textit{ChmMint} shown here),
so we only present one material for each type here.

\begin{figure}[t]
    \centering
    \includegraphics[width=0.99\linewidth]{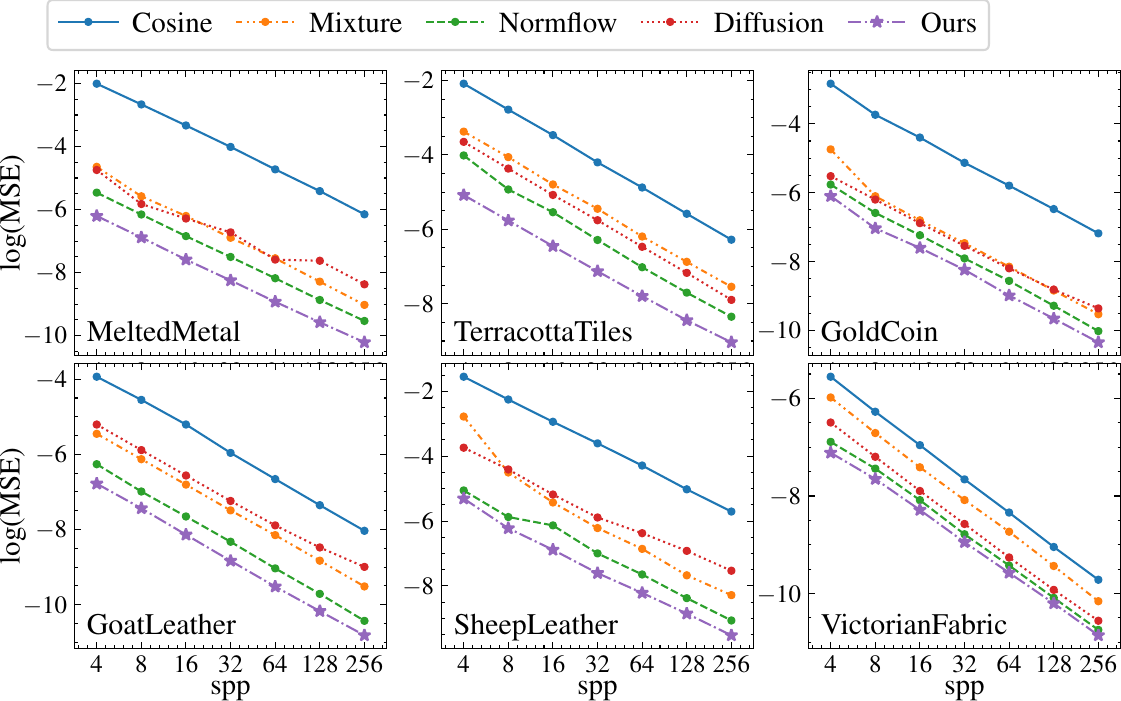}
    \caption{\textbf{Equal-spp convergence graph under constant light on the NeuSample materials.}
    The difference between each method is more noticeable than that in Fig.~12 of the paper.
    }
    \label{fig:6-sanity}
\end{figure}

\paragraph{Convergence graphs under constant light.}
On NeuSample materials,
\citeauthor{xu2023neusample}’s convergence graphs show more noticeable performance
gain because they are recorded on renderings with only BRDF sampling and constant light ($L_i(\mathbf{x},\bm{\omega}_i)\equiv1$).
We show the equal-spp convergence graph using the same setting in \cref{fig:6-sanity}.
With the incident radiance being constant,
only the BRDF term contributes to the variance,
such that each method gives more effective variance reduction than that in Fig.~12 of the paper.
This, however, does not change the relative performance between each method,
and our method still gives the best overall variance reduction.

\begin{figure}[t]
    \centering
    \includegraphics[width=0.9\linewidth]{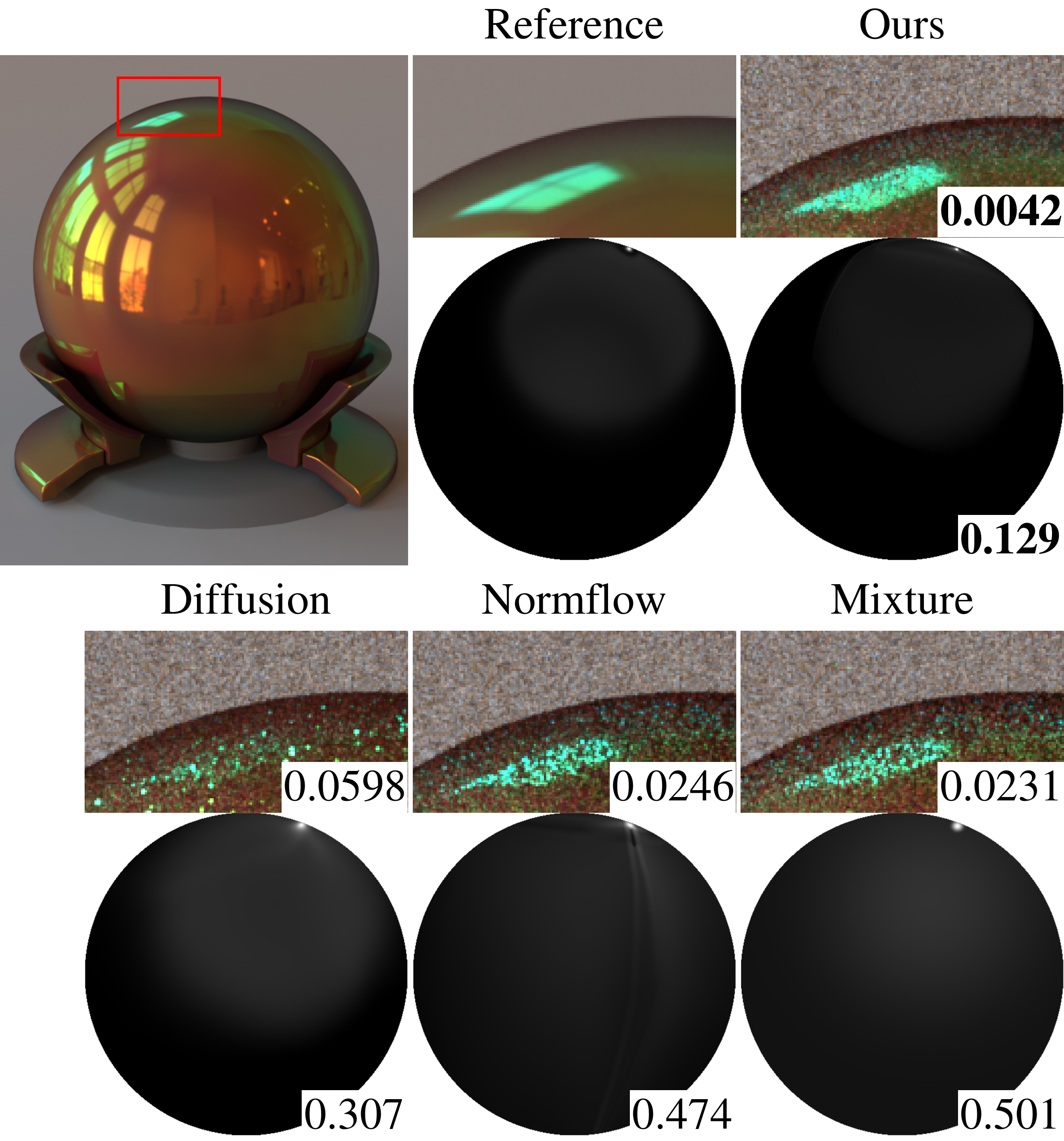}
    
    \caption{\textbf{Our method is robust to grazing angles.}
    The odd rows show image crops rendered using 4 spp with the numbers denoting the MSEs. The even rows show reconstructed pdfs at the grazing angle with the numbers denoting the KL-divergence.
    }
    \label{fig:6-grazing}
\end{figure}
\paragraph{Discussion of grazing angles.}
The BRDF sampling near grazing angles can be problematic~\cite{fu2024importance} which is demonstrated in \cref{fig:6-grazing}.
Here, the pdf reconstruction quality for \textit{AmberCitrine} is slightly worse then that in \cref{fig:6-plot}.
However, our method still maintains the best accuracy and gives the grazing angle renderings with the least variance.

\begin{figure*}[t]
    \centering
    \includegraphics[width=0.98\linewidth]{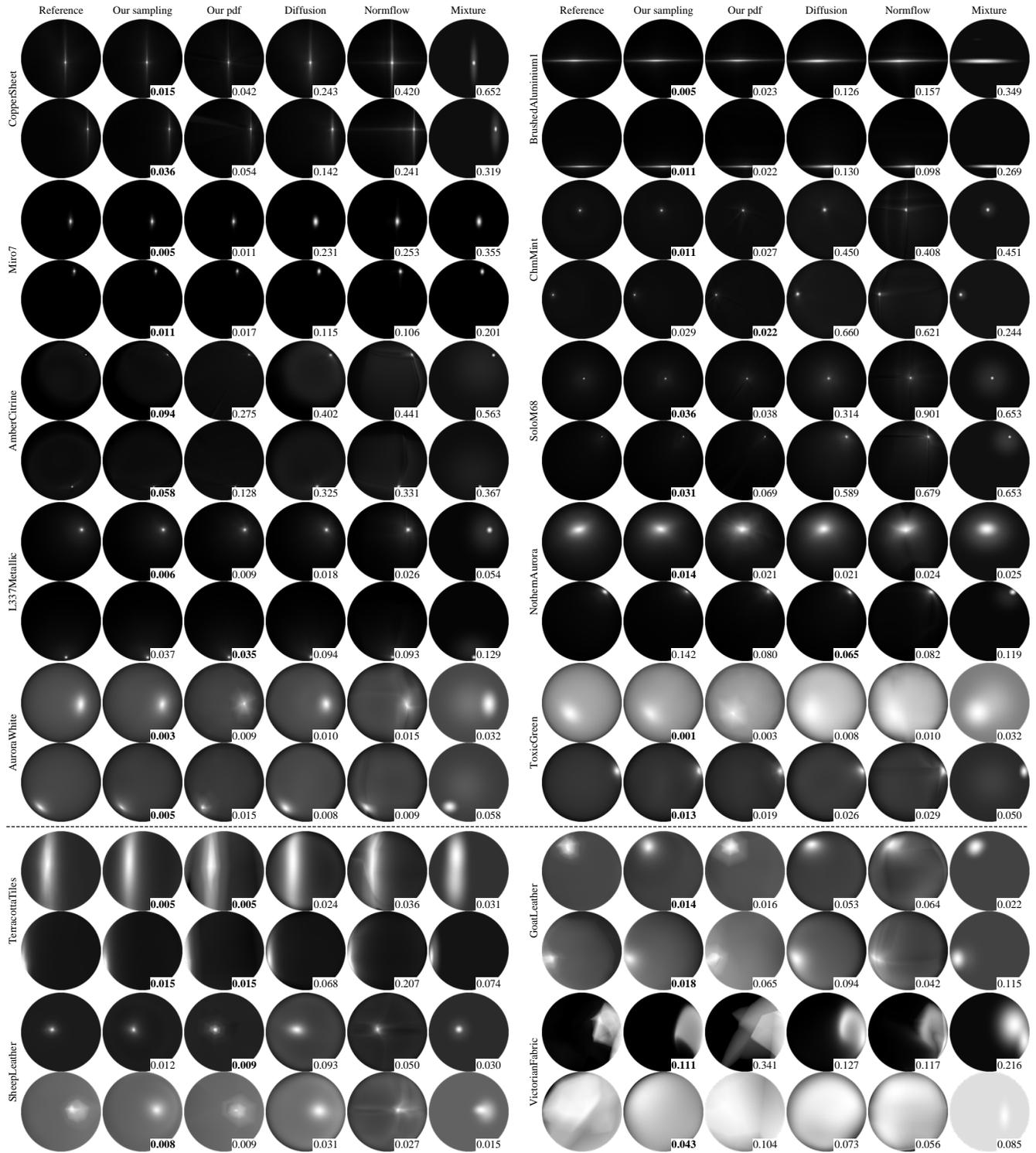}
    \caption{\textbf{Additional visualization of reconstructed pdfs}. Row 1-10 show RGL materials, and row 11-12 show NeuMIP materials.}
    \label{fig:6-plot}
\end{figure*}
\begin{figure*}[t]
    \centering
    \includegraphics[width=0.91\linewidth]{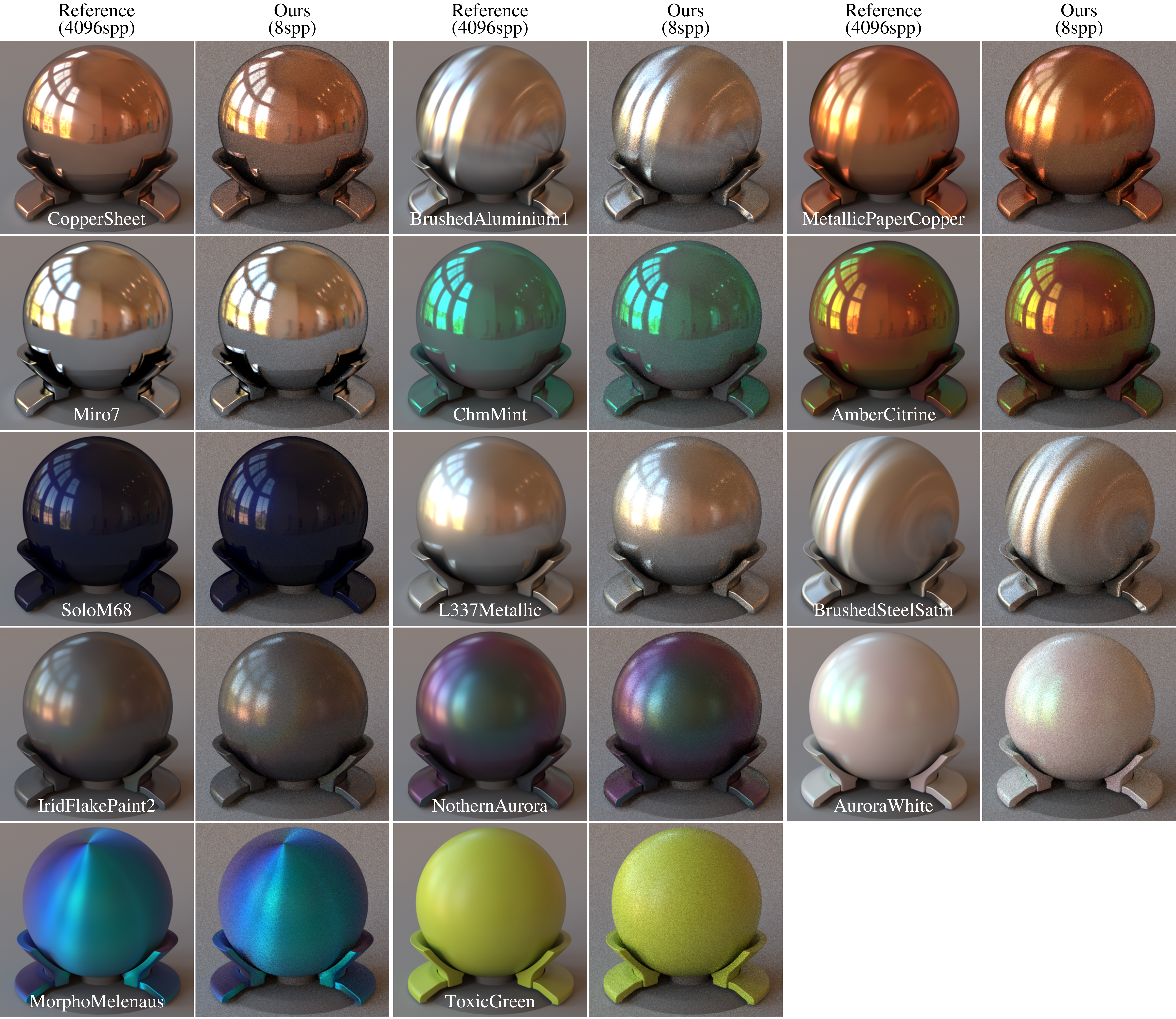}
    \caption{\textbf{Additional results on RGL materials.}}
    \label{fig:6-rendering1}
\end{figure*}
\begin{figure*}[t]
    \centering
    \includegraphics[width=0.91\linewidth]{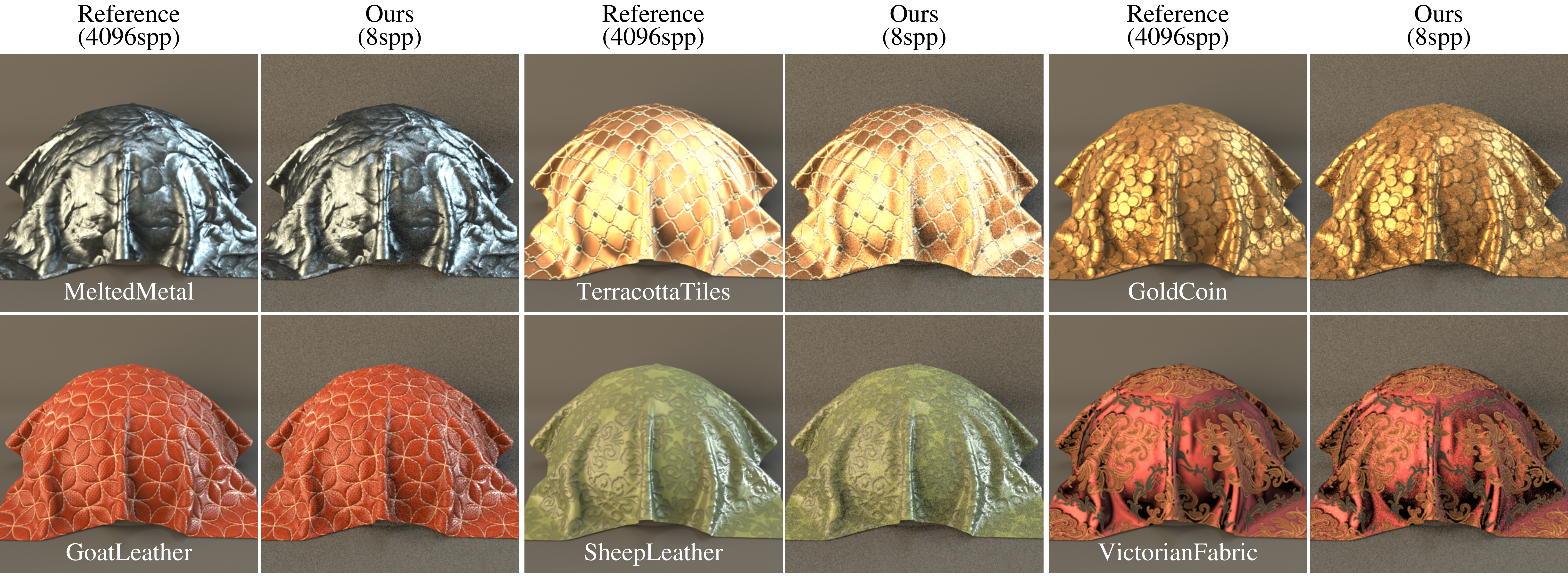}
    \caption{\textbf{Additional results on NeuSample materials.}}
    \label{fig:6-rendering2}
\end{figure*}
\end{appendix}
\end{document}